\documentclass[aps,groupedaddress]{revtex4}
\usepackage{epsfig}
\usepackage{graphicx}
\usepackage[T1]{fontenc}
\usepackage{ae}
\usepackage{color}
\usepackage[latin1]{inputenc}
\usepackage{amssymb,amsbsy,amsmath}
\usepackage{bbm}

\newtheorem{prop}{Proposition}

\begin{document}

%\begin{center}\today\end{center}

%%%%%%%%%%%%%%%%%%%%%%%%%%%%%%%%%%%%%%%%%%%%%%%%%%%%%%%%%%%%%%%%%%%%%%%%%%%%%

\title[Elliptical Rabi problem]
      {The Rabi problem with elliptical polarization}
\author{Heinz-J\"urgen Schmidt
}
\address{ Universit\"at Osnabr\"uck,
Fachbereich Physik,
 D - 49069 Osnabr\"uck, Germany}

%\tableofcontents

\begin{abstract}
We consider the solution of the equation of motion of a classical/quantum spin subject to a monochromatical, elliptically polarized external field.
The classical Rabi problem can be reduced to third order differential equations with polynomial coefficients and hence solved in terms
of power series in close analogy to the confluent Heun equation occurring for linear polarization. Application of Floquet theory
yields physically interesting quantities like the  quasienergy as a function of the problem's parameters
and expressions for the Bloch-Siegert shift of resonance frequencies. Various limit cases cases are thoroughly investigated.
\end{abstract}

\maketitle

%%%%%%%%%%%%%%%%%%%%%%%%%%%%%%%%%%%%%%%%%%%%%%%%%%%%%%%%%%%%%%%%%%%%%%%%%%%%%%%%%%%%%%%%%%%%%%%%%%%%%%%%%%%%%%%%%%%%%%%%%%%%%%%
\section{Introduction}\label{sec:I}
%%%%%%%%%%%%%%%%%%%%%%%%%%%%%%%%%%%%%%%%%%%%%%%%%%%%%%%%%%%%%%%%%%%%%%%%%%%%%%%%%%%%%%%%%%%%%%%%%%%%%%%%%%%%%%%%%%%%%%%%%%%%%%%

In recent years, theoretical and experimental evidence has shown that periodic driving
can be a key element for engineering exotic quantum mechanical states of matter,
such as time crystals and superconductors at room temperature \cite{OK19}, \cite{H16}, \cite{RL20}.
The renewed interest in Floquet engineering,
i.~e., the control of quantum systems by periodic driving,
is due to (a) the rapid development of laser and ultrashort spectroscopy techniques \cite{BAP15},
(b) the discovery and understanding of various ``quantum materials" that
exhibit interesting exotic properties \cite{RL19}, \cite{Fetal19} ,
and (c) the interaction with other emerging fields of physics such as programmable matter
\cite{TM91} and periodic thermodynamics \cite{Kohn01} - \cite{DSSH20}.

One of the simplest system to study periodic driving is a two level system (TLS) interacting
with a classical periodic radiation field. The special case of a constant magnetic field in, say,
$x$-direction plus a circularly polarized field in the $y-z$-plane was already solved more
than eight decades ago by I.~I.~Rabi \cite{R37} and can be found in many textbooks.
This case is referred to as RPC in the following.
Shortly thereafter, F.~Bloch and A.~Siegert \cite{BS40} considered the analogous problem of a
linearly polarized magnetic field orthogonal to the direction of the constant field (henceforth called RPL)
and proposed the so-called rotating wave approximation.
They also investigated the shift of the resonant frequencies due to
the approximation error of the rotating wave approximation,
since then called the "Bloch-Siegert shift."

In the following decades one noticed \cite{AT55}, \cite{S65} that
the underlying mathematical problem leads to the Floquet Theory
\cite{Floquet83}, which deals with linear differential matrix equations
with periodic coefficients \cite{YS75}, \cite{T12}.
Accordingly, analytical approximations for solutions were worked out,
which formed the basis for subsequent research. In particular, the
groundbreaking work of J.~H.~Shirley \cite{S65} is has been cited over 13,000 times.
Among the numerous applications of the theory of periodically driven
TLS are nuclear magnetic resonance \cite{R38}, ac-driven quantum dots \cite{WT10},
Josephson qubit circuits \cite{YN11}, and coherent destruction of tunnelling \cite{MZ16}.
On a theoretical level, the methods for solving the RPL and related problems have been gradually refined
and include power series approximations for Bloch-Siegert shifts \cite{HPS73}, \cite{AB74},
perturbation theory and/or various boundary cases \cite{LP92} -- \cite{WY07}
and the hybridized rotating wave approximation \cite{YLZ15}. Also, the inverse method yields
analytical solutions for certain periodically driven TLS \cite{GDG10} -- \cite{SNGM18}.

In the meantime also the RPL has been analytically solved \cite{ML07}, \cite{XH10}.
This solution is based on a transformation of the Schr\"odinger equation into a confluent Heun differential equation.
A similar approach was previously applied to the TLS subject to a magnetic pulse \cite{JR10}, \cite{JR10a}
and has been extended to other cases of physical interest \cite{IG14}, \cite{ISI15}.
In the special case of the RPL the analytical solution has been further elaborated to include time evolution over a full period
and explicit expressions for the quasienergy \cite{SSH20a}.

In this paper we will extend these results to the Rabi problem with elliptic polarization (RPE)
that is also of experimental interest, see \cite{Letal14}, \cite{Ketal15}.
Here we will approach the Floquet problem of the TLS via its well-known classical limit, see, e.~g., \cite{AW05}.
It has been shown that, for the particular problem of a TLS with periodic driving,
the classical limit is already equivalent to the quantum problem \cite{S18}, \cite{S20b}.
More precisely, to each periodic solution of the classical equation of motion there exists a Floquet solution
of the original Schr\"odinger equation that can be explicitly calculated via integrations. Especially, the quasienergy is essentially given by the action
integral over one period of the classical solution. This is reminiscent of the semiclassical Floquet theory developed in \cite{BH91}.

The motion of a classical spin vector ${\mathbf S}(\tau)$ in a monochromatical magnetic field with elliptic polarization and an orthogonal
constant component can be analyzed by following an approach analogous to that leading to the confluent Heun equation in \cite{ML07}, \cite{XH10}.
We differentiate the first order equation of motion twice and eliminate two components of the spin vector.
The resulting third order differential equation for the remaining component $x(\tau)$ can be transformed into a differential
equation with polynomial coefficients by the change from the dimensionless time variable $\tau$ to $u=\sin^2 \frac{\tau}{2}$.
The latter differential equation is solved by a power series in $u$ such that its coefficients satisfy a six terms recurrence relation.
The second component $y(\tau)$ can be treated in the same way whereas the third component $z(\tau)$ is obtained in a different way.
As in the RPL case the transformation from $\tau$ to $u$ is confined to the half period $0< \tau < \pi$ and,
and moreover, the resulting power series diverges for $u=1$ corresponding to $\tau=\pi$.
Hence it is necessary to reduce the full time evolution of the classical spin vector to the first quarter period. This is done
analogously to the procedure in \cite{SSH20a} utilizing the discrete symmetries of the polarization ellipse.

The structure of the paper is the following. In Section \ref{sec:Def} we present  the scenario of the classical Rabi problem
with elliptic polarization and its connection to the underlying Schr\"odinger equation. The above-mentioned
reduction of the time evolution to the first quarter period is made in Section \ref{sec:R}.
Already in the following Section \ref{sec:Q}, before solving the equation of motion,
it can be shown that the fully periodic monodromic matrix depends only on two parameters $r$ and $\alpha$,
which determine the quasienergy and the initial value of the periodic solution ${\mathbf S}(\tau)$, respectively.
The Fourier series of this solution necessarily have the structure of an even/odd $\cos$-series for $x(\tau)/y(\tau)$ and
an odd $\sin$-series for $z(\tau)$. The third order differential equations for $X(u(\tau))=x(\tau)$ and $Y(u(\tau))=y(\tau)$
and their power series solutions are derived in Section \ref{sec:H}. First consequences of this solution for the Fourier series coefficients
and the parameters $r$ and $\alpha$ are considered in Section \ref{sec:FS}. In order to check our results obtained so far we
consider, in Section \ref{sec:T}, an example of the time evolution with simple values of the parameters of the polarization ellipse
and two different initial values. On the one hand, we calculate the time evolution by using ten terms of the above-mentioned power series solutions
for the first quarter period and extend the result to the full period. One the other hand, we numerically calculate the time
evolution and find satisfactory agreement between both methods.

The quasienergy is discussed in more details in Section \ref{sec:QIII} with the emphasis on curves in parameter space where it
vanishes.
The resonance frequencies $\omega_{res}^{(n)}$ can be expressed in terms of power series in the variables $F$ and $G$ denoting the
semi-axes of the polarization ellipse and compared with known results for the limit cases of linear and circular polarization,
see Section \ref{sec:RES}.
The next Section \ref{sec:L} is devoted to the discussion of further limit cases along the lines of \cite{S18}.
In the adiabatic limit of vanishing driving frequency $\omega\rightarrow 0$
the spin vector follows the direction of the magnetic field, see Subsection \ref{sec:LOM}.
The corresponding quasienergy can be expressed through a complete elliptic integral of the second kind. The next two order
corrections proportional to $\omega^1$ and $\omega^2$ can be obtained recursively and yield a kind of asymptotic envelope
of a certain branch of the quasienergy as a function of $\omega$. In the next limit case of $F,G\rightarrow 0$ in Subsection \ref{sec:FG}
the solution ${\mathbf S}(t)$ and the quasienergy can be written in the form of a so-called Fourier-Taylor series.
This series is also of interest for the limit case of vanishing energy level splitting $\omega_0\rightarrow 0$ in Subsection \ref{sec:LOMO},
where it replaces the exact solution of the RPL for $\omega_0=0$, and allows analytical approximations for the further limit cases
$F\rightarrow 0$ and $F\rightarrow G$. An application concerning the work performed on a TLS by an elliptically polarized 
field is given in Section \ref{sec:WO}. We close with a summary and outlook in Section \ref{sec:SO}.

%%%%%%%%%%%%%%%%%%%%%%%%%%%%%%%%%%%%%%%%%%%%%%%%%%%%%%%%%%%%%%%%%%%%%%%%%%%%%%%%%%%%%%%%%%%%%%%%%%%%%%%%%%%%%%%%%%%%%%%%%%%%%%%
\section{The classical Rabi problem: General Definitions and results}\label{sec:Def}
%%%%%%%%%%%%%%%%%%%%%%%%%%%%%%%%%%%%%%%%%%%%%%%%%%%%%%%%%%%%%%%%%%%%%%%%%%%%%%%%%%%%%%%%%%%%%%%%%%%%%%%%%%%%%%%%%%%%%%%%%%%%%%%

We consider the Schr\"odinger equation
\begin{equation}\label{SE1}
 {\sf i}\hbar\, \frac{d}{dt}\Psi(t)=H(t)\,\Psi(t)
 \;,
\end{equation}
of a spin with quantum number $s=1/2$, $\Psi(t)={\Psi_1(t)\choose \Psi_2(t)}$ and a time-dependent, periodic
Hamiltonian
\begin{equation}\label{SE2}
 H(t) = \frac{\hbar}{2}\left( \omega_0\,\sigma_1+G\,\cos(\omega t)\,\sigma_2+F\,\sin(\omega t)\,\sigma_3\right)
 \;,
\end{equation}
where the $\sigma_i,\;i=1,2,3$ are the Pauli matrices
\begin{equation}\label{SE3}
 \sigma_1=\left(\begin{array}{cc}
                  0 & 1 \\
                  1 & 0
                \end{array} \right),\;
 \sigma_2=\left(\begin{array}{cc}
                  0 & -{\sf i} \\
                  {\sf i} & 0
                \end{array} \right),\;
 \sigma_3=\left(\begin{array}{cc}
                  1 & 0 \\
                  0 & -1
                \end{array} \right)\;.
\end{equation}
Hence $H(t)$ can be understood as a Zeeman term w.~r.~t.~a (dimensionless) magnetic field
\begin{equation}\label{SE4}
 {\mathbf H}(t)=\left(
 \begin{array}{c}
   \omega_0 \\
   G\,\cos\omega t \\
   F\,\sin\omega t
 \end{array}
 \right)
 \;.
\end{equation}
Alternatively, $\frac{\hbar}{2}\omega_0\,\sigma_1$ can be understood as the zero field Hamiltonian of a
two level system and (\ref{SE4}) without the constant component as a monochromatic,
elliptically polarized magnetic field.

Setting $\hbar=1$ and passing to a dimensionless time variable $\tau=\omega\,t$ we may rewrite (\ref{SE1}) in the form
\begin{equation}\label{SE5}
  {\sf i}\,\frac{d}{d\tau} {\psi_1(\tau)\choose \psi_2(\tau)}=
  {\textstyle\frac{1}{2}}
  \left(
  \begin{array}{cc}
    f\,\sin\tau & \nu-g\,\cos\tau \\
    \nu+g\,\cos\tau & -f\,\sin\tau
  \end{array}
    \right)\,
   {\psi_1(\tau)\choose \psi_2(\tau)}
   \;,
\end{equation}
where $G=g\,\omega$, $F=f\,\omega$  and $ \omega_0=\nu\,\omega$. The dimensionless period is always $T\,\omega=2\pi$.
Sometimes, we will denote the derivative w.~r.~t.~$\tau$ by an overdot $\frac{d}{d\tau}=\dot{}\;$.

Let
\begin{equation}\label{SE6}
 P(\tau)=\left|  {\psi_1(\tau)\choose \psi_2(\tau)} \right\rangle \left\langle  {\psi_1(\tau)\choose \psi_2(\tau)}\right|
\end{equation}
denote the one-dimensional time-dependent projector onto a solution of (\ref{SE5}) and
\begin{equation}\label{SE7}
  P(\tau) = {\textstyle\frac{1}{2}}\left({\mathbbm 1}+ x(\tau)\,\sigma_1+  y(\tau)\,\sigma_2+ z(\tau)\,\sigma_3\right)
\end{equation}
its expansion w.~r.~t.~the basis $({\mathbbm 1},\sigma_1,\sigma_2,\sigma_3)$ of Hermitean $2\times 2$-matrices.
It follows that the  vector ${\mathbf S}(\tau)=(x(\tau),y(\tau),z(\tau))^\top$
satisfies the classical equation of motion
\begin{equation}\label{D1}
 \frac{d}{d\tau}{\mathbf S}(\tau) = {\mathbf h}(\tau)\times {\mathbf S}(\tau)
 \;,
\end{equation}
and hence ${\mathbf S}(\tau)$ can be viewed as a classical spin vector (not necessarily normalized).
Moreover,
\begin{equation}\label{D2}
 {\mathbf h}(\tau)=\left(
 \begin{array}{c}
   h_1 \\
  h_2 \\
   h_3
 \end{array}
 \right)
 =\left(
 \begin{array}{c}
   \nu \\
   g\,\cos\tau \\
   f\,\sin\tau
 \end{array}
 \right)
\end{equation}
denotes the dimensionless magnetic field vector (\ref{SE4}) written as a function of $\tau$.

Conversely, to each solution of (\ref{D1}) one obtains the corresponding solution of (\ref{SE5})
up to a time-dependent phase that can be obtained by an integration, see \cite{S18} for the details.

The coefficients of the Taylor series w.~r.~t.~$\tau$ of $x(\tau),y(\tau)$ and $z(\tau)$ can be recursively
determined by using (\ref{D1}) and the initial values $x(0),y(0)$ and $z(0))$. Note that
$h_1$ and $h_2$ are even functions of $\tau$ and that $h_3$ is an odd one. Hence there exist
special solutions of (\ref{D1}) such that $x(\tau)$ and $y(\tau)$ are even functions of $\tau$ and
$z(\tau)$ is an odd one, symbolically:
\begin{equation}\label{Deo1}
 {\mathbf S}(\tau)=\left(
 \begin{array}{c}
   \mbox{even} \\
  \mbox{even}\\
  \mbox{odd}
 \end{array}
 \right)
 \;.
\end{equation}
In fact, this is consistent with (\ref{D1}) and (\ref{D2}) since
\begin{equation}\label{Deo2}
 \frac{d}{d \tau}{\mathbf S}(\tau)=\left(
 \begin{array}{c}
   \mbox{odd} \\
  \mbox{odd}\\
  \mbox{even}
 \end{array}
 \right)
 \;,
\end{equation}
and
\begin{equation}\label{Deo3}
{\mathbf h}\times {\mathbf S}=\left(
 \begin{array}{c}
   \mbox{even} \\
  \mbox{even}\\
  \mbox{odd}
 \end{array}
 \right)
 \times
 \left(
 \begin{array}{c}
   \mbox{even} \\
  \mbox{even}\\
  \mbox{odd}
 \end{array}
 \right)=
 \left(
 \begin{array}{c}
   \mbox{odd} \\
  \mbox{odd}\\
  \mbox{even}
 \end{array}
 \right)
 \;,
\end{equation}
and can be proven by induction over the degree of the Taylor series coefficients of $ {\mathbf S}(\tau)$
using the necessary initial condition $z(0)=0$.

Analogously, there exist solutions ${\mathbf S}(\tau)$ of type
\begin{equation}\label{Deo4}
 {\mathbf S}(\tau)=\left(
 \begin{array}{c}
   \mbox{odd} \\
  \mbox{odd}\\
  \mbox{even}
 \end{array}
 \right)
 \;.
\end{equation}
satisfying $x(0)=y(0)=0$. We will state these results in the following form:
%%%%%%%%%%%%%%%%%%%%%%%%%%%%%%%%%%%%%%%%%%%%%%%%%%%%%%%%%%%%%%%%%%%%%%%%%%%%%%%%%%%%%%%%%%%%%%%%%%%%%%%%%%%%%%%%%%%%%%%%%%%%%%%%%%%%%%%%%%%%%%%%%
\begin{prop}\label{Peo}
\begin{enumerate}
  \item The solution ${\mathbf S}(\tau)$ of (\ref{D1}) is of type (\ref{Deo1}) iff $z(0)=0$.
  \item Analogously, the solution ${\mathbf S}(\tau)$ of (\ref{D1}) is of type (\ref{Deo4}) iff $x(0)=y(0)=0$.
\end{enumerate}
\end{prop}
%%%%%%%%%%%%%%%%%%%%%%%%%%%%%%%%%%%%%%%%%%%%%%%%%%%%%%%%%%%%%%%%%%%%%%%%%%%%%%%%%%%%%%%%%%%%%%%%%%%%%%%%%%%%%%%%%%%%%%%%%%%%%%%%%%%%%%%%%%%%%%%%%%%%
For general initial conditions the solution ${\mathbf S}(\tau)$ of (\ref{D1}) will be of mixed type.\\

Next, let ${\mathbf S}^{(i)}(\tau),\;i=1,2,3,$ denote the three solutions of (\ref{D1}) with initial conditions
${\mathbf S}^{(i)}_j(\tau_0)=\delta_{ij}$ and $R(\tau,\tau_0)$ be the $3\times 3$-matrix with columns ${\mathbf S}^{(i)}(\tau)$.
Since the  ${\mathbf S}^{(i)}(\tau)$ are mutually orthogonal and and right-handed for $\tau=\tau_0$ this holds for all
$\tau\in{\mathbbm R}$ and hence $R(\tau,\tau_0)\in SO(3)$.
It satisfies the differential equation
\begin{equation}\label{D3}
  \frac{d}{d\tau}R(\tau,\tau_0)=H(\tau)\, R(\tau,\tau_0)
  \;,
\end{equation}
with initial condition
\begin{equation}\label{D4}
 R(\tau_0,\tau_0)={\mathbbm 1}
 \;.
\end{equation}
Here $H(\tau)\in so(3)$ is the real anti-symmetric $3\times 3$-matrix corresponding to ${\mathbf h}(\tau)$, i.~e.~,
\begin{equation}\label{D5}
 H(\tau)=
 \left(
 \begin{array}{ccc}
  0 & -f\,\sin\tau & g\,\cos\tau \\
   f\,\sin\tau & 0 & -\nu \\
   -g\,\cos\tau & \nu & 0
 \end{array}
 \right)
 \;.
\end{equation}
The differential equation (\ref{D3}) with initial condition (\ref{D4}) has a unique solution $R(\tau,\tau_0)$ for all $\tau,\,\tau_0\in{\mathbbm R}$, see, e.~g.,
theorem $3.9$ in \cite{T12}.
Obviously, this implies the composition law
\begin{equation}\label{DRcomp}
 R(\tau_2,\tau_0)=  R(\tau_2,\tau_1)\, R(\tau_1,\tau_0)
 \;,
\end{equation}
and hence
\begin{equation}\label{DRinv}
 R(\tau_2,\tau_1)^{-1}=  R(\tau_1,\tau_2)
 \;,
\end{equation}
for all $\tau_0,\tau_1, \tau_2\in{\mathbbm R}$.

Usually we will set $\tau_0=0$.
The matrix $H(\tau)$ is obviously $2\pi$-periodic.
Hence we may apply Floquet theory to the classical equation of motion (\ref{D1}).
The monodromy matrix $R(2\pi,0)$ has the eigenvalues $\{1,\exp\left(\pm {\sf i}\rho\}\right)$ which leads to
the corresponding classical quasienergy (or Floquet exponent) of the form
\begin{equation}\label{D6}
 \epsilon^{(cl)}=0,\pm \frac{\rho}{2\pi}
 \;,
\end{equation}
uniquely defined up to integer multiples (note that effectively $\omega=1$ in our approach).

The connection to the quasienergy $\epsilon^{(qu)}$ of the underlying spin $s=\frac{1}{2}$ Schr\"odinger equation
(\ref{SE5})
can be given in two ways. Either we may utilize the fact that the classical Rabi problem can
be understood as the ``lift" of the spin $s=\frac{1}{2}$ problem to spin $s=1$. Then Eq.~(38) of
\cite{S18} implies
\begin{equation}\label{D7}
  \epsilon^{(cl)}=2\, m\,\epsilon^{(qu)},\quad \mbox{where } m=-1,0,1
  \;.
\end{equation}
Taking into account the mentioned ambiguity of $\epsilon^{(cl)}$ this means that we have two
possibilities: Either $\epsilon^{(qu)}=\pm \frac{1}{2} \epsilon^{(cl)}$ or
$\epsilon^{(qu)}=\frac{1}{2}(1\pm \epsilon^{(cl)})$. Since we have, modulo integers, only two
values for $\epsilon^{(qu)}$ these two possibilities are generally exclusive. One way to decide
between the two possibilities would be to utilize the well-known quasienergies for the RPC,
that agree with the case $\epsilon^{(qu)}=\frac{1}{2}(1\pm \epsilon^{(cl)})$, and to argue with
continuity.

Another way to obtain $\epsilon^{(qu)}$ would be to follow the prescription given in \cite{S18} and write
\begin{equation}\label{D8}
 \epsilon^{(qu)}=\overline{\frac{1}{2}\left(h_1+\frac{h_2\,y+h_3\,z}{1+z} \right)}
 \;,
\end{equation}
where the overline indicates the time average over one period of a $2\pi$-periodic solution
${\mathbf S}(\tau)$ of (\ref{D1}). An equivalent expression,
that is manifestly invariant under rotations, is given by
\begin{equation}\label{D8a}
 \epsilon^{(cl)}=
 \textstyle{
 \overline{{\mathbf h}\cdot {\mathbf S} -
 \frac{{\mathbf S}\cdot\left(\dot{\mathbf S}\times \ddot{\mathbf S} \right)}{\dot{\mathbf S}\cdot \dot{\mathbf S}}}
 }
 \;,
\end{equation}
see Eq.~(46) in \cite{S20b}.
Periodic solutions of (\ref{D1}) can be found by using the
initial value ${\mathbf S}(0)={\mathbf r}$, where ${\mathbf r}$ is the normalized eigenvector
of $R(2\pi,0)$ corresponding to the eigenvalue $1$, see also \cite{S20b}.

Of course, both ways, (\ref{D7}) and (\ref{D8}), to obtain $\epsilon^{(qu)}$ agree within the usual ambiguity modulo integers.
This will be explicitly checked in Section \ref{sec:QIII} for the case of circular polarization.

%%%%%%%%%%%%%%%%%%%%%%%%%%%%%%%%%%%%%%%%%%%%%%%%%%%%%%%%%%%%%%%%%%%%%%%%%%%%%%%%%%%%%%%%%%%%%%%%%%%%%%%%%%%%%%%%%%%%%%%%%%%%%%%
\section{Reduction to the first quarter period}\label{sec:R}
%%%%%%%%%%%%%%%%%%%%%%%%%%%%%%%%%%%%%%%%%%%%%%%%%%%%%%%%%%%%%%%%%%%%%%%%%%%%%%%%%%%%%%%%%%%%%%%%%%%%%%%%%%%%%%%%%%%%%%%%%%%%%%%

Due to the discrete symmetries of the polarization ellipse
it is possible to reduce the time evolution of the classical
spin to the first quarter period $\tau\in [0,\frac{\pi}{2}]$. This is similar to the corresponding considerations in \cite{SSH20a}.
Let $T^{(i)},\,i=1,2,3$ denote the involutory diagonal $3\times 3$-matrices with entries $T^{(i)}_{jk}=(-1)^{\delta_{ij}}\,\delta_{jk}$, for example,
\begin{equation}\label{R1}
 T^{(1)}=\left(
 \begin{array}{rrr}
  -1&0 & 0\\
   0 & 1 &0 \\
   0 & 0 & 1
 \end{array}
 \right)
 \;,
\end{equation}
and $T^{(ij)}\equiv T^{(i)}\,T^{(j)}$, for example,
\begin{equation}\label{R2}
 T^{(13)}=\left(
 \begin{array}{rrr}
  -1&0 & 0\\
   0 & 1 &0 \\
   0 & 0 & -1
 \end{array}
 \right)
 \;.
\end{equation}

First we will formulate a proposition that allows us to reduce the time evolution
for the classical spin from the full period to the first half period $\tau\in[0,\pi]$.
%%%%%%%%%%%%%%%%%%%%%%%%%%%%%%%%%%%%%%%%%%%%%%%%%%%%%%%%%%%%%%%%%%%%%%%%%%%%%%%%%%%%%%%%%%%%%%%%%%%%%%%%%%%%%%%%%%%%%%%%%%%%%%%%%%%%%%%%%%%%%%%%%%%%%%%%%%%%%%%%%
\begin{prop}\label{prop1}
 \begin{equation}\label{R3}
  R(\pi+\tau,0)=T^{(1)}\,R(\tau,0)\,T^{(1)}\,R(\pi,0)
 \end{equation}
 for all $\tau\in{\mathbbm R}$.
\end{prop}
%%%%%%%%%%%%%%%%%%%%%%%%%%%%%%%%%%%%%%%%%%%%%%%%%%%%%%%%%%%%%%%%%%%%%%%%%%%%%%%%%%%%%%%%%%%%%%%%%%%%%%%%%%%%%%%%%%%%%%%%%%%%%%%%%%%%%%%%%%%%%%%%%%%%%%%%%%%%%%%%%
\noindent Proof:\\
Let $\tilde{R}(\tau)\equiv T^{(1)}\,R(\pi+\tau,\pi)\,T^{(1)}$
such that $\tilde{R}(0)=T^{(1)}\,R(\pi,\pi)\,T^{(1)}={\mathbbm 1}$.
It satisfies the differential equation
\begin{eqnarray}
\label{R4a}
  \frac{d}{d\tau} \tilde{R}(\tau)&=& T^{(1)}\,\left( \frac{d}{d\tau} R(\pi+\tau,\pi)\right)\,T^{(1)} \\
  \label{R4b}
   &\stackrel{(\ref{D3})}{=}& T^{(1)}\,\left(H(\pi+\tau)\, R(\pi+\tau,\pi)\right)\,T^{(1)}\\
   \label{R4c}
   &=& \left(T^{(1)}\,H(\pi+\tau)\,T^{(1)}\right)\left(T^{(1)}\, R(\pi+\tau,\pi)\,T^{(1)}\right)\\
    \label{R4d}
   &=& H(\tau)\,\left(T^{(1)}\, R(\pi+\tau,\pi)\,T^{(1)}\right)\\
   \label{R4e}
   &=& H(\tau)\, \tilde{R}(\tau)
   \;.
\end{eqnarray}
In (\ref{R4d}) we have used that $\sin( \pi+\tau)= -\sin\tau$, $\cos(\pi+\tau)=-\cos\tau$ and hence
\begin{equation}\label{R5}
 T^{(1)}\,H(\pi+\tau)\,T^{(1)}=
\left(
 \begin{array}{rrr}
  -1&0 & 0\\
   0 & 1 &0 \\
   0 & 0 & 1
 \end{array}
 \right)\;
 \left(
 \begin{array}{ccc}
  0 & f\,\sin\tau &- g\,\cos\tau \\
   -f\,\sin\tau & 0 & -\nu \\
   g\,\cos\tau & \nu & 0
 \end{array}
 \right)
 \;
 \left(
 \begin{array}{rrr}
  -1&0 & 0\\
   0 & 1 &0 \\
   0 & 0 & 1
 \end{array}
 \right)
 \end{equation}

 \begin{equation}\label{R6}
 =
 \left(
 \begin{array}{ccc}
  0 & -f\,\sin\tau & g\,\cos\tau \\
   f\,\sin\tau & 0 & -\nu \\
   -g\,\cos\tau & \nu & 0
 \end{array}
 \right)
 =H(\tau)
 \;.
\end{equation}
It follows that $\tilde{R}(\tau)$ satisfies the same differential equation and initial condition as $R(\tau,0)$
and hence\\
$T^{(1)}\,R(\pi+\tau,\pi)\,T^{(1)}=\tilde{R}(\tau)=R(\tau,0)$. Consequently,
\begin{equation}\label{R7}
 R(\pi+\tau,0)\stackrel{(\ref{DRcomp})}{=}R(\pi+\tau,\pi)\,R(\pi,0)=T^{(1)}\,R(\tau,0)\,T^{(1)}\,R(\pi,0)
 \;,
\end{equation}
which completes the proof of the proposition.  \hfill$\Box$\\

Setting $\tau=\pi$ in (\ref{R3}) gives
\begin{equation}\label{R7a}
  R(2\pi,0)=T^{(1)}\,R(\pi,0)\,T^{(1)}\,R(\pi,0)=\left(T^{(1)}\,R(\pi,0)\right)^2
  \;.
 \end{equation}

Next we show how to further reduce the time evolution to the first quarter period $\tau\in [0,\frac{\pi}{2}]$.
%%%%%%%%%%%%%%%%%%%%%%%%%%%%%%%%%%%%%%%%%%%%%%%%%%%%%%%%%%%%%%%%%%%%%%%%%%%%%
\begin{prop}\label{prop2}
 \begin{equation}\label{R8}
  R(\pi-\tau,0)=T^{(13)}\,R(\tau,0)\,T^{(13)}\,R(\pi,0)
 \end{equation}
 for all $\tau\ge 0$.
\end{prop}
%%%%%%%%%%%%%%%%%%%%%%%%%%%%%%%%%%%%%%%%%%%%%%%%%%%%%%%%%%%%%%%%%%%%%%%%%%%%%
\noindent Proof:
The proof is similar to that of proposition \ref{prop1} except that an additional time reflection is involved.
Let $\tilde{R}(\tau)\equiv T^{(13)}\,R(\pi-\tau,\pi)\,T^{(13)}$ such that
$\tilde{R}(0)=T^{(13)}\,R(\pi,\pi)\,T^{(13)}={\mathbbm 1}$.
It satisfies the differential equation
\begin{eqnarray}
\label{R9a}
  \frac{d}{d\tau} \tilde{R}(\tau)&=& T^{(13)}\,\left( \frac{d}{d\tau} R(\pi-\tau,\pi)\right)\,T^{(13)} \\
  \label{R9b}
   &\stackrel{(\ref{D3})}{=}& T^{(13)}\,\left(-H(\pi-\tau)\, R(\pi-\tau,\pi)\right)\,T^{(13)}\\
   \label{R9c}
   &=& \left(T^{(13)}\,\left(-H(\pi-\tau)\,T^{(13)}\right)\right)\left(T^{(13)}\, R(\pi-\tau,\pi)\,T^{(13)}\right)\\
    \label{R9d}
   &=& H(\tau)\,\left(T^{(13)}\, R(\pi-\tau,\pi)\,T^{(13)}\right)\\
   \label{R9e}
   &=& H(\tau)\, \tilde{R}(\tau)
   \;.
\end{eqnarray}
In (\ref{R9d}) we have used that $\sin (\pi-\tau)= \sin\tau$, $\cos (\pi-\tau)= -\cos\tau$ and hence
\begin{equation}\label{R10}
 T^{(13)}\,(-H(\pi-\tau))\,T^{(13)}=
\left(
 \begin{array}{rrr}
  -1&0 & 0\\
   0 & 1 &0 \\
   0 & 0 & -1
 \end{array}
 \right)\;
 \left(
 \begin{array}{ccc}
  0 & f\,\sin\tau & g\,\cos\tau \\
   -f\,\sin\tau & 0 & \nu \\
   -g\,\cos\tau & -\nu & 0
 \end{array}
 \right)
 \;
 \left(
 \begin{array}{rrr}
  -1&0 & 0\\
   0 & 1 &0 \\
   0 & 0 & -1
 \end{array}
 \right)
 \end{equation}

 \begin{equation}\label{R11}
 =
 \left(
 \begin{array}{ccc}
  0 & -f\,\sin\tau & g\,\cos\tau \\
   f\,\sin\tau & 0 & -\nu \\
   -g\,\cos\tau & \nu & 0
 \end{array}
 \right)
 =H(\tau)
 \;.
\end{equation}
It follows that $\tilde{R}(\tau)$ satisfies the same differential equation and initial condition as $R(\tau,0)$
and hence
\begin{equation}\label{R11a}
T^{(13)}\,R(\pi-\tau,\pi)\,T^{(13)}=\tilde{R}(\tau)=R(\tau,0)
 \;.
\end{equation}
Consequently,
\begin{equation}\label{R12}
 R(\pi-\tau,0)\stackrel{(\ref{DRcomp})}{=} R(\pi-\tau,\pi)\, R(\pi,0) = T^{(13)}\, R(\tau,0)\,T^{(13)}\,R(\pi,0)
 \;,
\end{equation}
which completes the proof of the proposition.  \hfill$\Box$\\

Setting $\tau=\pi$ in (\ref{R11a}) implies
\begin{equation}\label{R12a}
T^{(13)}\,R(0,\pi)\,T^{(13)}=R(\pi,0)
  \;,
\end{equation}
and hence
\begin{equation}\label{R12b}
R(0,\pi)\stackrel{(\ref{DRinv})}{=}R(\pi,0)^\top=T^{(13)}\,R(\pi,0)\,T^{(13)}
  \;.
\end{equation}
Moreover, if we set $\tau=\frac{\pi}{2}$ in (\ref{R8}) we obtain
\begin{equation}\label{R13}
  R({\textstyle\frac{\pi}{2},0})=T^{(13)}\, R({\textstyle\frac{\pi}{2},0})\,T^{(13)}\,R(\pi,0)
  \;,
\end{equation}
and hence, solving for $ R(\pi,0)$,
\begin{equation}\label{R14}
  R(\pi,0)=T^{(13)}\, R({\textstyle\frac{\pi}{2},0})^\top\,T^{(13)}\, R({\textstyle\frac{\pi}{2},0})
  \;.
\end{equation}
Thus (\ref{R8}) can be re-written as
\begin{equation}\label{R15}
  R(\pi-\tau,0)=T^{(13)}\,R(\tau,0)\, R({\textstyle\frac{\pi}{2},0})^\top\,T^{(13)}\, R({\textstyle\frac{\pi}{2},0})
  \;,
 \end{equation}
 and hence the evolution data for $\tau\in [\frac{\pi}{2},\pi]$ can be completely written in terms
 of those for $\tau\in[0,\frac{\pi}{2}]$. Together with (\ref{R3}) this shows that the complete time
 evolution can be reduced to that in the first quarter period.

%%%%%%%%%%%%%%%%%%%%%%%%%%%%%%%%%%%%%%%%%%%%%%%%%%%%%%%%%%%%%%%%%%%%%%%%%%%%%%%%%%%%%%%%%%%%%%%%%%%%%%%%%%%%%%%%%%%%%%%%%%%%%%%
\section{Fourier series and quasienergy: Preliminary results}\label{sec:Q}
%%%%%%%%%%%%%%%%%%%%%%%%%%%%%%%%%%%%%%%%%%%%%%%%%%%%%%%%%%%%%%%%%%%%%%%%%%%%%%%%%%%%%%%%%%%%%%%%%%%%%%%%%%%%%%%%%%%%%%%%%%%%%%%

First we will re-derive (\ref{R12b}) under more general assumptions.
\begin{prop}\label{prop3}
Let $ R\in SO(3)$ and $T\in O(3)$ be such that $T^2={\mathbbm 1}$ and hence $T^\top=T$. Define $\tilde{R}\in SO(3)$ by
 \begin{equation}\label{Q1}
   \tilde{R}\equiv T\,R^\top\,T\,R
   \;,
  \end{equation}
  then
  \begin{equation}\label{Q2}
   \tilde{R}^\top = T\,\tilde{R}\,T
  \end{equation}
  holds.
\end{prop}
Proof:
$\tilde{R}^\top =\left(T\,R^\top\,T\,R\right)^\top= R^\top\,T\,R\,T=T^2\,R^\top\,T\,R\,T
=T\,\left( T\,R^\top\,T\,R\right)\,T=T\,\tilde{R}\,T$. \hfill$\Box$\\

Let us specialize to the case $T=T^{(13)}$, then (\ref{Q2}) is equivalent to the following three equations:
\begin{equation}\label{Q3}
 \tilde{R}_{12}=-\tilde{R}_{21},\quad \tilde{R}_{13}=\tilde{R}_{31},\quad \tilde{R}_{23}=-\tilde{R}_{32}\;.
\end{equation}
A general rotational matrix $\tilde{R}\in SO(3)$ can be determined by three real parameters; by the three equations (\ref{Q3})
the number of parameters can be reduced to two:
\begin{prop}\label{prop4}
Every rotational matrix $\tilde{R}\in SO(3)$ satisfying (\ref{Q3}) will be of the form
\begin{equation}\label{Q4}
\tilde{R}=\left(
\begin{array}{ccc}
 r^2+\left(1-r^2\right) \cos (2 \alpha ) & \left(1-r^2\right) \sin (2 \alpha ) & 2 r
   \sqrt{1-r^2} \sin (\alpha ) \\
 -\left(1-r^2\right) \sin (2 \alpha ) &\left(1-r^2\right) \cos (2 \alpha )-r^2 & 2 r
   \sqrt{1-r^2} \cos (\alpha ) \\
 2 r \sqrt{1-r^2} \sin (\alpha ) & -2 r \sqrt{1-r^2} \cos (\alpha ) & 1-2 r^2 \\
\end{array}
\right)\;,
\end{equation}
where $r\in [0,1]$ and $\alpha\in [0,2\pi]$.
\end{prop}
Proof:\\
Obviously, the $3^{rd}$ column $\tilde{R}_3$  of $\tilde{R}$ according to (\ref{Q4}) is the most general form of a unit vector.
The $2^{nd}$ column $\tilde{R}_2$ must be a unit vector orthogonal to $\tilde{R}_3$
with a given component $\tilde{R}_{3,2}=-\tilde{R}_{2,3}=-2 r \sqrt{1-r^2} \cos (\alpha )$.
If $r>0$ there are only two possibilities for $\tilde{R}_2$: the first one is given by
(\ref{Q4}) and the second one is
${\mathbf b}=\left(-r^2 \sin (2 \alpha ), -r^2\cos (2 \alpha )-r^2+1,-2 r \sqrt{1-r^2} \cos(\alpha )\right)^\top$.
The $1^{st}$ column of $\tilde{R}$ is uniquely given by $\tilde{R}_1=\tilde{R}_2\times \tilde{R}_3$,
but $\tilde{R}_2={\mathbf b}$ does not yield a matrix satisfying (\ref{Q3}) and hence has to be excluded.

We have still to consider the case $r=0$ such that $\tilde{R}_3=(0,0,1)^\top$.
Then the representation (\ref{Q4}) reduces to
\begin{equation}\label{Q5}
 \left(
\begin{array}{ccc}
 \cos (2 \alpha ) & \sin (2 \alpha ) & 0 \\
 -\sin (2 \alpha ) & \cos (2 \alpha ) & 0 \\
 0 & 0 & 1 \\
\end{array}
\right)
\;,
\end{equation}
which is obviously the most general case satisfying (\ref{Q3}) and $\tilde{R}_3=(0,0,1)^\top$.  \hfill$\Box$\\

Recall that the ``half period monodromy matrix" $R(\pi,0)$ satisfies (\ref{R12b}), hence, according to Prop.~\ref{prop3}, also (\ref{Q3}) and,
by virtue of Prop.~\ref{prop4}, must be of the form (\ref{Q4}). In the case of linear polarization ($g=0$)
this result also follows from the form of the half period monodromy matrix $U(\pi,0)$ of the corresponding
Schr\"odinger equation, see equation (30) in \cite{SSH20a}, where the parameters $r$ and $\alpha$ have the same
meaning as in this paper. Using (\ref{R7a}) we can immediately derive the form of the full period monodromy matrix
\begin{equation}\label{Q6}
 R(2\pi,0)=\left(
\begin{array}{ccc}
 \left(1-2 r^2\right)^2+4 r^2 \left(1-r^2\right) \cos (2 \alpha ) & 4 r^2
   \left(1-r^2\right) \sin (2 \alpha ) & 4 r \sqrt{1-r^2} \left(2 r^2-1\right) \sin
   (\alpha ) \\
 4 r^2 \left(1-r^2\right) \sin (2 \alpha ) & 4 \left(r^2-1\right) r^2 \cos (2 \alpha
   )+\left(1-2 r^2\right)^2 & 4 r \left(1-2 r^2\right) \sqrt{1-r^2} \cos (\alpha ) \\
 4 r \left(1-2 r^2\right) \sqrt{1-r^2} \sin (\alpha ) & 4 r \sqrt{1-r^2} \left(2
   r^2-1\right) \cos (\alpha ) & 8 r^4-8 r^2+1 \\
\end{array}
\right).
\end{equation}

It will be instructive to sketch another derivation of (\ref{Q6}). To this end we state
without proof that the monodromy matrix of the Schr\"odinger equation (\ref{SE5}) will assume the form
\begin{equation}\label{QQ1}
 U= U(2\pi,0)=\left(
 \begin{array}{cc}
  1-2 r^2 & 2{\sf i}r\sqrt{1-r^2}e^{-{\sf i}\alpha} \\
   2{\sf i}r\sqrt{1-r^2}e^{{\sf i}\alpha} & 1-2 r^2
 \end{array}
 \right)\;,
\end{equation}
completely analogous to eq.~(33) of \cite{SSH20a}.
$U$ has the eigenvalues $\exp (\pm 2 \,{\sf i}\,\arcsin r)$ with respective eigenvectors
$\left(e^{\pm{\sf i}\alpha},1\right)^\top$. Then the corresponding monodromy matrix $\rho$
of the classical RPE is given by the equation
\begin{equation}\label{QQ2}
  U\,\sigma_i\,U^\ast = \sum_{j=1}^{3}\rho_{j,i}\,\sigma_j
  \;,
\end{equation}
where the $\sigma_i$ are the Pauli matrices (\ref{SE3}).
It is easy to check that the so defined matrix $\rho$ coincides with $R(2\pi,0)$ given by (\ref{Q6}).

Like $R(\pi,0)$ also $R(2\pi,0)$ depends only on two parameters $\alpha$ and $r$ and satisfies a similar equation
that characterizes the corresponding two-dimensional submanifold of $SO(3)$, to wit,
\begin{equation}\label{QQ3}
R(2\pi,0)^\top = T^{(3)}\,R(2\pi,0)\,T^{(3)}
\;.
\end{equation}
This equation can be proven either directly by checking (\ref{Q6}) or by applying (\ref{R7a}) and (\ref{R12b}).

According to the general theory  \cite{S18} the eigenvalues of $R(2\pi,0)$  that are generally of the form $(1,\exp (\pm{\sf i}\rho))$
yield the quasienergies $\epsilon_\pm^{(qu)}$ of the underlying Schr\"odinger equation for spin $s=\frac{1}{2}$ via
\begin{equation}\label{Q7}
\exp \pm{\sf i}\rho=\exp\left( 4\,\pi\,{\sf i}\,\epsilon_\pm^{(qu)}\right)
 \;.
\end{equation}
As in \cite{SSH20a} it follows that
\begin{equation}\label{Q8}
  \epsilon_\pm^{(qu)}=\pm\frac{1}{4 \pi }\arg \left(1+8 r^4-8 r^2+4 {\sf i}r \left(1-2 r^2\right) \sqrt{1-r^2}\right)
  =\pm\frac{1}{\pi} \arcsin r
  \;.
\end{equation}
The eigenvector ${\mathbf r}$ corresponding to the real eigenvalue $1$ of $R(2\pi,0)$ is
\begin{equation}\label{Q9}
 {\mathbf r}=\left( \begin{array}{c}
                      \cos\alpha \\
                      \sin\alpha \\
                      0
                    \end{array}
             \right)
             \;.
\end{equation}
Choosing ${\mathbf r}$ as the initial value ${\mathbf r}={\mathbf S}(0)$ for the time evolution (\ref{D1})
yields a $2\pi$-periodic solution. Any other unit vector in the plane $P$ orthogonal to ${\mathbf r}$
will, in general, not return to its initial value after the time $\tau=2\pi$ but will be rotated in the plane $P$
by the angle $2\pi\epsilon^{(cl)}$. This endows the parameters $r$ and $\alpha$ occurring in
(\ref{Q4}) and (\ref{Q6}) with a geometrical and dynamical meaning.

Another remarkable result follows from $R(\pi,0)$ being of the form (\ref{Q4}):
\begin{equation}\label{Q10}
 R(\pi,0)\,\left( \begin{array}{c}
                      \cos\alpha \\
                      \sin\alpha \\
                      0
                    \end{array}
             \right)=
             \left( \begin{array}{c}
                      \cos\alpha \\
                      -\sin\alpha \\
                      0
                    \end{array}
             \right)
             \;,
\end{equation}
which means that for the initial value ${\mathbf S}(0)={\mathbf r}$ the half period time evolution
is equivalent to a reflection at the $x-z$-plane. This has further consequences for the Fourier series
of the $2\pi$-periodic functions $x(\tau),\,y(\tau),$ and $z(\tau)$ with initial values
$x(0)=\cos\alpha,\;y(0)=\sin\alpha$ and $z(0)=0$. Since $x(\tau)$ and $y(\tau)$ will be even functions of $\tau$
and $z(\tau)$ will be an odd one, see Proposition \ref{Peo}, we can write their Fourier series in the form
\begin{eqnarray}
\label{Q11a}
  x(\tau) &=& \sum_{\mu=0}^\infty x_\mu\,\cos(\mu\tau),\\
  \label{Q11b}
   y(\tau) &=& \sum_{\mu=0}^\infty y_\mu\,\cos(\mu\tau),\\
  \label{Q11c}
   x(\tau) &=& \sum_{\mu=1}^\infty z_\mu\,\sin(\mu\tau)
   \;.
\end{eqnarray}
Now consider the sequence of linear mappings
\begin{equation}\label{Q12}
 {\mathbf r}= \left( \begin{array}{c}
                      \cos\alpha \\
                      \sin\alpha \\
                      0
                    \end{array}
             \right)
  \stackrel{R(\pi,0)}{\longrightarrow}
  \left( \begin{array}{c}
                      \cos\alpha \\
                      -\sin\alpha \\
                      0
                    \end{array}
             \right)
      \stackrel{T^{(1)}}{\longrightarrow}
      \left( \begin{array}{c}
                     -\cos\alpha \\
                      -\sin\alpha \\
                      0
                    \end{array}
             \right)
      \stackrel{R(\tau,0)}{\longrightarrow}
      \left( \begin{array}{c}
                     -x(\tau) \\
                      -y(\tau) \\
                      -z(\tau)
                    \end{array}
             \right)
      \stackrel{T^{(1)}}{\longrightarrow}
      \left( \begin{array}{c}
                     x(\tau) \\
                      -y(\tau) \\
                      -z(\tau)
                    \end{array}
             \right)
              \stackrel{(\ref{R3})}{=}
              R(\pi+\tau,0)\,{\mathbf r}
              =
               \left( \begin{array}{c}
                     x(\pi+\tau) \\
                      y(\pi+\tau) \\
                     z(\pi+\tau)
                    \end{array}
             \right)
              \;.
  \end{equation}
From this we conclude
\begin{equation}\label{Q13}
 x(\pi+\tau)=\sum_{\mu=0}^\infty x_\mu\,\cos(\mu(\pi+\tau))=
 \sum_{\mu\,\mbox{ \scriptsize even}} x_\mu\,\cos(\mu\tau)\,
-\sum_{\mu\,\mbox{ \scriptsize odd}}x_\mu\,\cos(\mu\tau)
=x(\tau)= \sum_{\mu\,\mbox{ \scriptsize even}} x_\mu\,\cos(\mu\tau)\,
+\sum_{\mu\,\mbox{ \scriptsize odd}}x_\mu\,\cos(\mu\tau)
\;.
\end{equation}
Hence the odd terms of the $\cos$-series must vanish and $x(\tau)$ is an even $\cos$-series.
Similarly we conclude from (\ref{Q12}) that $y(\tau)$ is an odd $\cos$-series and $z(\tau)$ an odd $\sin$-series.
Summarizing, we have proven the following
\begin{prop}\label{prop5}
The components of the $2\pi$-periodic solution  ${\mathbf S}(\tau)$ of (\ref{D1}) with initial values
${\mathbf S}(0)={\mathbf r}$ according to (\ref{Q9})
have the Fourier series
\begin{eqnarray}
\label{Q14a}
  x(\tau) &=& \sum_{\mu \text{ even}}x_\mu\,\cos(\mu\tau), \\
  \label{Q14b}
  y(\tau) &=& \sum_{\mu\text{ odd}}y_\mu\,\cos(\mu\tau), \\
  \label{Q14c}
  z(\tau) &=& \sum_{\mu\text{ odd}}z_\mu\,\sin(\mu\tau)
   \;.
\end{eqnarray}
 In particular, the time averages of $y(\tau)$ and $z(\tau)$ over one period vanish.
\end{prop}

%%%%%%%%%%%%%%%%%%%%%%%%%%%%%%%%%%%%%%%%%%%%%%%%%%%%%%%%%%%%%%%%%%%%%%%%%%%%%%%%%%%%%%%%%%%%%%%%%%%%%%%%%%%%%%%%%%%%%%%%%%%%%%%
\section{Third order differential equations for single spin components}\label{sec:H}
%%%%%%%%%%%%%%%%%%%%%%%%%%%%%%%%%%%%%%%%%%%%%%%%%%%%%%%%%%%%%%%%%%%%%%%%%%%%%%%%%%%%%%%%%%%%%%%%%%%%%%%%%%%%%%%%%%%%%%%%%%%%%%%

We consider again  (\ref{D1}) and its higher derivatives that read
\begin{equation}\label{H1}
\frac{d}{d\tau} {\mathbf S}= \left(\begin{array}{c}
          \dot{x} \\
          \dot{y} \\
          \dot{z}
        \end{array}
  \right)=
  \left(
\begin{array}{c}
 g z \cos \tau-f y \sin \tau \\
 f x \sin \tau-\nu  z \\
 \nu  y-g x \cos \tau \\
\end{array}
\right)
\;,
\end{equation}

\begin{equation}\label{H2}
\frac{d^2}{d\tau^2} {\mathbf S}= \left(\begin{array}{c}
          \ddot{x} \\
          \ddot{y} \\
          \ddot{z}
        \end{array}
  \right)=
  \left(
\begin{array}{c}
 -\sin \tau \left(f^2 x \sin \tau+z (g-f \nu )\right)+y \cos \tau (g \nu -f)-g^2 x \cos^2\tau \\
  \cos \tau(x (f+g \nu )+f g z \sin\tau)-y \left(f^2 \sin ^2\tau+\nu   ^2\right) \\
   \sin \tau (x (f \nu +g)+f g y \cos \tau)-z \left(g^2 \cos ^2\tau +\nu    ^2\right) \\
\end{array}
\right)
\;,
\end{equation}

\begin{equation}\label{H3}
\frac{d^3}{d\tau^3} {\mathbf S}= \left(\begin{array}{c}
          \dddot{x} \\
          \dddot{y} \\
          \dddot{z}
        \end{array}
  \right)=: x\,  {\mathbf S}^{(3)}_1+ y\,  {\mathbf S}^{(3)}_2+ z\,  {\mathbf S}^{(3)}_3
\;,
\end{equation}
with
\begin{equation}\label{H3a}
 {\mathbf S}^{(3)}_1=
 \left(
\begin{array}{c}
 -3 \left(f^2-g^2\right) \sin \tau \,\cos\tau \\
 -\sin\tau \left(f^3 \sin ^2\tau+f g^2 \cos ^2\tau+f \nu ^2+f+g \nu
   \right) \\
 \cos \tau \left(f^2 g \sin ^2\tau+f \nu +g^3 \cos ^2\tau+g \nu
   ^2+g\right) \\
\end{array}
\right)
 \;,
\end{equation}

\begin{equation}\label{H3b}
 {\mathbf S}^{(3)}_2=
\left(
\begin{array}{c}
 \sin \tau \left(f^3 \sin ^2\tau+f g^2 \cos ^2\tau+f \nu ^2+f-2 g \nu
   \right) \\
 -3 f^2 \sin\tau\, \cos \tau \\
 -f \sin ^2\tau (f \nu +2 g)-g \cos ^2\tau (g \nu -f)-\nu ^3 \\
\end{array}
\right)
 \;,
\end{equation}
and
\begin{equation}\label{H3c}
 {\mathbf S}^{(3)}_3=
 \left(
\begin{array}{c}
 -\cos \tau \left(f^2 g \sin ^2\tau-2 f \nu +g^3 \cos ^2\tau+g \nu
   ^2+g\right) \\
 f \sin ^2\tau (f \nu -g)+g \cos ^2\tau (2 f+g \nu )+\nu ^3 \\
 3 g^2 \sin \tau\, \cos\tau \\
\end{array}
\right)
 \;.
\end{equation}

It is obvious that $\dot{x}$ and $\ddot{x}$ depend linearly on  $y$ and $z$ and that this dependence
can be inverted to express $y$ and $z$ in terms of $x$, $\dot{x}$ and $\ddot{x}$. Inserting this
result into $\dddot{x}$ yields a third order linear differential equation for $x(\tau)$ where the coefficients
are trigonometric functions of $\tau$.

Similarly, we can obtain third order differential equations for $y(\tau)$ and $z(\tau)$.
For the preparation of the next step we make the restriction to solutions of (\ref{H1})
such that $x(\tau)$ and $y(\tau)$ are even functions of $\tau$ whereas $z(\tau)$ is an odd one,
according to Prop.~\ref{Peo}.
In this way we could obtain two solutions ${\mathbf S}^{(1)}$ and ${\mathbf S}^{(2)}$
with different initial conditions for $x(\tau)$ and $y(\tau)$ and the initial condition $z(0)=0$,
the latter being a consequence of the restriction to odd functions $z(\tau)$.
The third solution ${\mathbf S}^{(3)}$ with  $x(\tau)$ and $y(\tau)$ odd and $z(\tau)$ even is then uniquely determined
by ${\mathbf S}^{(1)}$ and ${\mathbf S}^{(2)}$. For example, if ${\mathbf S}^{(1)}$ and ${\mathbf S}^{(2)}$ are chosen to be
orthogonal for $\tau=0$ then they will be orthogonal for all $\tau$ and ${\mathbf S}^{(3)}$ is just the vector product
of ${\mathbf S}^{(1)}$ and ${\mathbf S}^{(2)}$.

Following \cite{XH10} we will consider a transformation $\tau\mapsto u$ of the independent variable
such that the coefficients of the transformed differential equations become rational functions of $u$.
This transformation will be chosen as
\begin{equation}\label{H4}
  u(\tau)=\sin^2\frac{\tau}{2}=\frac{1}{2}\left(1-\cos\tau\right)
  \;,
\end{equation}
the same as in \cite{XH10}, and maps the half period $\tau\in[0,\pi]$ bijectively onto $u\in[0,1]$.
Since (\ref{H4}) defines an even function of $\tau$ the corresponding transformation is only appropriate
for the even functions $x(\tau)$ and $y(\tau)$. Their transforms will be denoted by $X(u)$ and $Y(u)$
such that
\begin{equation}\label{H5}
X(u(\tau))=x(\tau),\quad \mbox{and } Y(u(\tau))=y(\tau)
\mbox{ for } \tau\in[0,\pi]
\;.
\end{equation}
The remaining function $z(\tau)$ has to be calculated differently, e.~g.,
by using that the length of ${\mathbf S}(\tau)$ is conserved under time evolution according to (\ref{D1}).
This gives the result
\begin{equation}\label{H6}
  z(\tau)=\pm\sqrt{x(0)^2+y(0)^2-x(\tau)^2-y(\tau)^2}
  \;,
\end{equation}
where $z(0)=0$ has been used, and the sign has to be chosen in such a way that $z(\tau)$ remains a smooth function
in the neighbourhood of its zeros. An alternative procedure would be possible if $x(\tau)$ and $y(\tau)$ can be
written as Fourier series (maybe only locally valid for $\tau\in[0,\pi/2]$). Then $z(\tau)$ could be
obtained by a direct integration of $\dot{z}(\tau)= \nu  y(\tau)-g x(\tau)\, \cos (\tau )$.
This last procedure will be applied in Section \ref{sec:T}.

We come back to the differential equation for $X(u)$ and write it with polynomial coefficients $p_n(u)$ in the form
\begin{equation}\label{H7}
  0=\sum_{n=0}^3 p_n(u) X^{(n)}(u)
  \;.
\end{equation}
The coefficients are the following ones:
\begin{eqnarray}
\label{H8a}
 p_3(u) &=& u \left(1-u)(4 f^2 \nu  (u-1) u+f g-g^2 \nu  (1-2 u)^2\right)\\
 \label{H8b}
 p_2(u) &=& -\frac{1}{2} (2 u-1) \left(4 f^2 \nu  (u-1) u+3 f g+g^2 \nu  (-4 (u-1)   u-3)\right) \\
 \nonumber
 p_1(u) &=& -16 f^4 \nu  (u-1)^2 u^2-4 f^3 g (u-1) u+4 f^2 \nu  (u-1) u \left(2 g^2 (1-2u)^2+\nu ^2\right)\\
 \label{H8c}
&& +f g^3 (1-2 u)^2+3 f g \nu ^2-g^2 \nu  \left(g^2 (1-2    u)^4+\nu ^2 (1-2 u)^2+2\right) \\
\label{H8d}
  p_0(u) &=& -2 (2 u-1) (f-g) (f+g) \left(4 f^2 \nu  (u-1) u+3 f g-g^2 \nu  (1-2 u)^2\right).
\end{eqnarray}
The singular points of the differential equation are the zeros of $p_3(u)$. Except the points
$u=0$ and $u=1$ that occur also for the confluent Heun equation, see \cite{XH10} and \cite{SSH20a},
we have an additional pair of singular points, real or complex ones, depending on the parameters $f,g$ and $\nu$.
The obvious ansatz to obtain a physically relevant solution of (\ref{H7}) is a power series
\begin{equation}\label{H8e}
  X(u)=\sum_{n=0}^\infty \xi_n\,u^n
\end{equation}
at the singular point given by $u=0$. We have not investigated its radius of convergence, but it is
clear that the series diverges at least for the second singular point $u=1$, which has been our motivation
to restrict the application of (\ref{H8e}) to $|u|\le{\textstyle\frac{1}{2}}$ corresponding to the first quarter period $\tau\in[0,\pi/2]$.
In contrast to \cite{XH10} we need only one real solution and can neglect further solutions of the fundamental system.
However, due to the degree three of the differential equation and the additional singular points we need
a six-term recurrence relation for the coefficients of the power series.

We will not give the details of the recurrence relation
but rather sketch how to obtain it by means of computer-algebraic aids.
We take a finite part $\sum_{n=m-3}^{m+2} \xi_n\,u^n$ of the power series and insert it into the
differential equation (\ref{H7}). The result is expanded into a $u$-polynomial and the coefficient of $u^m$
is set to $0$. It has been checked that only the above considered finite part of the power series influences
this coefficient. Thus we obtain a six-term recurrence relation of the form
\begin{equation}\label{H9}
\xi_{m+2}=\sum_{i=m-3}^{m+1}a_i\,\xi_i
\;,
\end{equation}
where the $a_i$ have been determined as rational functions of $f,g,\nu$,
but they are too complicated to be presented here.

The next problem is that we need the first five coefficients of $X(u)=\sum_n \xi_n\,u^n$
to get the next coefficients using the recursion relation. Since the original equation (\ref{D1})
is of $1^{st}$ order we have only two undetermined initial values $x(0)$ and $y(0)$, taking into account that $z(0)=0$.
To solve this problem we have compared the first terms of the $\tau$-power series of $x(\tau)$ and $X(u(\tau))$,
using the differential equation (\ref{D1}), and thereby determined $\xi_0,\ldots,\xi_4$ as functions of $x(0)$ and $y(0)$.
This also compensates the enlargement of the solution space by passing from a $1^{st}$ order differential equation
to a $3^{rd}$ order one. To give an impression of the kind of results we display the first three coefficients:
\begin{eqnarray}
\label{H10a}
  \xi_0 &=& x(0) \\
  \label{H10b}
  \xi_1 &=& -2 \left(y(0) (f-g \nu )+g^2 x(0)\right) \\
  \label{H10c}
  \xi_2 &=& \frac{1}{3} \left(2 x(0) \left(-3 f^2-2 f g \nu +g^2 \left(g^2+\nu
   ^2+3\right)\right)+2 y(0) \left(f \left(g^2+3 \nu ^2\right)-g \nu
   \left(g^2+\nu ^2+2\right)\right)\right)
   \;.
\end{eqnarray}
Obviously, $\xi_n$ is a linear function of $x(0)$ and $y(0)$ that can be written as
\begin{equation}\label{H10d}
  \xi_n=\xi_n^{(x)}\,x(0)+\xi_n^{(y)}\,y(0)
  \;.
\end{equation}

After these preparations it is, in principle, possible to calculate any finite number of power series coefficients
$\xi_n$ as a function of the physical parameters $f,g$ and $\nu$ and the initial values $x(0)$ and $y(0)$, although
the expressions become more and more intricate, and finally to obtain a truncated approximation of $X(u(\tau))$.
For a comparison to a numerical solution of (\ref{D1}) see Section \ref{sec:T}.\\

Analogous considerations apply for the case of the solution $y(\tau)=Y(u(\tau))$. This time we obtain a
differential equation of the form
\begin{equation}\label{H11}
  0=\sum_{n=0}^3 q_n(u) Y^{(n)}(u)
  \;,
\end{equation}
where
\begin{eqnarray}
\label{H12a}
 q_3(u) &=& (u-1) u (2 u-1)^3 \left(f^2 g+f \nu +g \nu ^2\right) \left(4 f^2 g (u-1) u-f \nu -g \nu ^2\right)\\
 \label{H12b}
 q_2(u) &=& \frac{1}{2} (1-2 u)^2 \left(f^2 g+f \nu +g \nu ^2\right) \left(4 f^2 g (u-1)u+f \nu  (-8 (u-1) u-3)+g \nu ^2 (-8 (u-1) u-3)\right) \\
 \nonumber
 q_1(u) &=&(1-2 u)^2 (2 u-1) \left(f^2 g+f \nu +g \nu^2\right) \left(16 f^4 g (u-1)^2
   u^2-4 f^3 \nu  (u-1) u-4 f^2 g (u-1) u \left(g^2 (1-2 u)^2 +2 \nu^2\right) \right. \\
 \label{H12c}
&& \left. +f \nu  \left(3 g^2 (1-2 u)^2+\nu ^2\right)+g^3 \nu ^2 (1-2 u)^2+g \nu ^4\right)\\
\nonumber
q_0(u)&=&2 (1-2 u)^2 \left(f^2 g+f \nu +g \nu ^2\right) \left(4 f^4 g (u-1) u+f^3 \nu
   (-8 (u-1) u-3)+f^2 g \nu ^2 (4 (u-1) u-1)-f \nu ^3-g \nu ^4\right).\\
   \label{H12d}
   &&
\end{eqnarray}
The zeros of $q_3(u)$ yield five singular points.
The power series solution ansatz
\begin{equation}\label{H12e}
  Y(u)=\sum_{n=0}^\infty \eta_n\,u^n
\end{equation}
leads to a $9$-term recursion relation
and the first $8$ coefficients are again determined by calculating the corresponding $t$-power series coefficients.
We show the first three ones.
\begin{eqnarray}
\label{H13a}
  \eta_0 &=& y(0) \\
  \label{H13b}
  \eta_1 &=& 2 x(0) (f+g \nu )-2 \nu ^2 y(0) \\
  \label{H13c}
  \eta_2 &=& \frac{1}{3} \left(2 y(0) \left(-3 f^2+2 f g \nu +\nu ^2 \left(g^2+\nu
   ^2-1\right)\right)-2 x(0) \left(f \left(3 g^2+\nu ^2\right)+g \nu
   \left(g^2+\nu ^2\right)\right)\right)
   \;.
\end{eqnarray}
Analogously to (\ref{H10d}), $\eta_n$ is a linear function of $x(0)$ and $y(0)$ that can be written as
\begin{equation}\label{H13d}
  \eta_n=\eta_n^{(x)}\,x(0)+\eta_n^{(y)}\,y(0)
  \;.
\end{equation}
The further details are too intricate to be displayed here, but, in principle, the procedure is completely analogous to
the power series solution of the confluent Heun equation investigated in \cite{SSH20a}.

%%%%%%%%%%%%%%%%%%%%%%%%%%%%%%%%%%%%%%%%%%%%%%%%%%%%%%%%%%%%%%%%%%%%%%%%%%%%%%%%%%%%%%%%%%%%%%%%%%%%%%%%%%%%%%%%%%%%%%%%%%%%%%%
\section{Fourier series and quasienergy: Results based on the power series solutions}\label{sec:FS}
%%%%%%%%%%%%%%%%%%%%%%%%%%%%%%%%%%%%%%%%%%%%%%%%%%%%%%%%%%%%%%%%%%%%%%%%%%%%%%%%%%%%%%%%%%%%%%%%%%%%%%%%%%%%%%%%%%%%%%%%%%%%%%%

It is clear that $u^n=\sin^{2n}\frac{\tau}{2}=\left(\frac{1}{2}\left(1-\cos\tau\right)\right)^n$ is a finite Fourier series
including only $\cos$-terms.
It explicitly reads
\begin{equation}\label{FS1}
 \sin^{2n}\frac{\tau}{2}=\frac{(2 n-1){!!}}{2^n n!}+\sum _{\mu =1}^n \frac{(2 n-1){!!} (1-\mu +n)_{\mu } (-1)^{\mu } }{2^{n-1} (\mu +n)!}\,\cos (\mu  \tau)
 \;,
\end{equation}
where $(a)_\mu=a(a+1)\ldots(a+\mu-1)$ denotes the Pochhammer symbol. Inserting (\ref{FS1}) into the power series (\ref{H8e}) and (\ref{H12e})
for $x(\tau)$ and $y(\tau)$ yields Fourier series representations valid within the convergence radius of the power series.
This does {\it not} mean that $x(\tau)$ and $y(\tau)$ are generally periodic functions but only that they locally,
within the respective domains of convergence,  coincide with periodic
functions. We may explicitly write down the corresponding Fourier coefficients
of
\begin{eqnarray}
\label{FSfourx}
  x(\tau) &=&\sum_{\mu=0}^\infty x_\mu\,\cos(\mu\tau)\;, \\
  \label{FSfoury}
  y(\tau) &=& \sum_{\mu=0}^\infty y_\mu\,\cos(\mu\tau)
  \;,
\end{eqnarray}
to wit,
\begin{eqnarray}
\label{FS2a}
x_\mu &=&
\left\{ \begin{array}{r@{\quad}l}
	\sum_{n=0}^\infty \frac{ (2 n-1)!!}{2^n n!}\,\xi_n &:\quad\mu=0, \\
&\\
	\sum _{n=\mu}^{\infty } \frac{(2 n-1){!!} (1-\mu+n)_\mu (-1)^\mu}{2^{n-1}   (\mu+n)!}\,\xi_n & :\quad \mu>0 ,
	\end{array} \right.\\
\label{FS2b}
y_\mu &=&
\left\{ \begin{array}{r@{\quad}l}
	\sum_{n=0}^\infty \frac{ (2 n-1)!!}{2^n n!}\,\eta_n &:\quad\mu=0, \\
&\\
	\sum _{n=\mu}^{\infty } \frac{(2 n-1){!!} (1-\mu+n)_\mu (-1)^\mu}{2^{n-1}   (\mu+n)!}\,\eta_n &:\quad \mu>0 .
	\end{array} \right.
\end{eqnarray}
Recall that the  $\xi_n$ and $\eta_n$ are the coefficients of the power series (\ref{H8e}) and (\ref{H12e}) to be determined by
means of recurrence relations.

The case of $z(\tau)$ is a bit more complicated. Using the above local Fourier series representation of $x(\tau)$ and $y(\tau)$
we may directly solve the differential equation
\begin{equation}\label{FSfourz}
  \dot{z}(\tau)=\nu\, y(\tau)-g\,\cos\tau\, x(\tau)
  \;,
\end{equation}
since the r.~h.~s.~of (\ref{FSfourz}) is again a $\cos$-series. In general, there will be a non-vanishing constant term $z_0$
at the r.~h.~s.~of  (\ref{FSfourz}) that generates a corresponding part $z_0\,\tau$ of $z(\tau)$ taking into account that $z(0)=0$.

The complete result is the following:
\begin{eqnarray}
\label{FS2d}
  z(\tau) &=&z_0\,\tau +\sum_{\mu=1}^{\infty} z_\mu\,\sin(\mu\,\tau)\;,\\
  \label{FS2e}
 z_\mu &=& \left\{ \begin{array}{r@{\quad:\quad}l}	\nu\,y_0-\frac{g}{2}\,x_1&\mu=0, \\
	 \nu\,y_1-g\,x_0-\frac{g}{2}\,x_2& \mu=1 ,\\
     \frac{1}{\mu}\left(\nu\,y_\mu-\frac{g}{2}\,\left( x_{\mu-1}+x_{\mu+1}\right)\right)& \mu>1.
	\end{array} \right.
 \end{eqnarray}

The expressions (\ref{FS2a}) and (\ref{FS2b}) for the Fourier coefficients still depend, via $\xi_n$ and $\eta_n$, on the
initial conditions $x(0)$ and $y(0)$. In the special case of $x(0)=\cos\alpha,\;y(0)=\sin\alpha$
according to (\ref{Q9})
the solutions $x(\tau)$ and $y(\tau)$ will be $2\pi$-periodic functions and hence, according to proposition \ref{prop5},
can be written as even resp.~odd $\cos$-series valid for all $\tau\in{\mathbbm R}$.
In particular,
\begin{equation}\label{FS3}
y_0=0= \sum_{n=0}^\infty \frac{ (2 n-1)!!}{2^n n!}\,\eta_n=\sum_{n=0}^\infty \frac{ (2 n-1)!!}{2^n n!}\,(\eta_n^{(x)}x(0)+\eta_n^{(y)}y(0))
=\sum_{n=0}^\infty \frac{ (2 n-1)!!}{2^n n!}\,(\eta_n^{(x)}\cos\alpha+\eta_n^{(y)}\sin\alpha)
\;.
\end{equation}
This equation can be solved for the auxiliary parameter $\alpha$:
\begin{equation}\label{FS4}
 \alpha=-\arctan\left(\frac{\sum_{n=0}^\infty \frac{ (2 n-1)!!}{2^n n!}\,\eta_n^{(x)}
 }{\sum_{n=0}^\infty \frac{ (2 n-1)!!}{2^n n!}\,\eta_n^{(y)}}
 \right)
 \;,
\end{equation}
if the numerator and denominator of this fraction do not vanish simultaneously.
This solution is only determined modulo $\pi$ in accordance with the fact that $x(0)=-\cos\alpha,\;y(0)=-\sin\alpha$
also gives a periodic solution.

The determination of the second auxiliary parameter $r$ is more involved.
We consider the following procedure that does not presuppose the determination of $\alpha$.
First we calculate the quarter period monodromy matrix $R(\frac{\pi}{2},0)$ by means of the local Fourier series
representation considered above. From this we obtain $R(\pi,0)$ via (\ref{R14}) and finally $r$ by
\begin{equation}\label{FS5}
 r=\pm\sqrt{\frac{1-R(\pi,0)_{3,3}}{2}}
 \,.
\end{equation}
The latter holds since $R(\pi,0)$ is of the form (\ref{Q4}). It will be instructive to give some more details.

First consider $R(\frac{\pi}{2},0)_{1,1}=x(\frac{\pi}{2})$, where $x(\tau)$ has the initial values $x(0)=1,\;y(0)=z(0)=0$.
It follows that
\begin{equation}\label{FS6}
 x\left({\textstyle\frac{\pi}{2}}\right)=\sum_{\mu=0}^{\infty}x_\mu \cos\left(\mu{\textstyle\frac{\pi}{2}}\right)=\sum_{\mu=0,4,\ldots}x_\mu-\sum_{\mu=2,6,\ldots}x_\mu
 \;,
\end{equation}
because the only non-vanishing terms are $\cos(\mu\frac{\pi}{2})=1$ for $\mu$ being an integer multiple of $4$
and  $\cos(\mu\frac{\pi}{2})=-1$ for even $\mu$ such that $\mu/2$ is odd. Recall that the Fourier coefficients $x_\mu$
have to be determined via (\ref{FS2a}) where the $\xi_n$ have to be chosen as $\xi_n^{(x)}$ according to (\ref{H10d})
and the above initial values.

The procedure for the calculation of $R(\frac{\pi}{2},0)_{2,1}=y(\frac{\pi}{2})$ is completely analogous.
For $R(\frac{\pi}{2},0)_{3,1}=z(\frac{\pi}{2})$ we employ (\ref{FS2d}) and (\ref{FS2e}) as well as
\begin{equation}\label{FS7}
z\left({\textstyle\frac{\pi}{2}}\right)=z_0\,{\textstyle\frac{\pi}{2}}+\sum_{\mu=1,5,\ldots}z_\mu-\sum_{\mu=3,7,\ldots}z_\mu
\;.
\end{equation}

For the $2^{nd}$ column of  $R(\frac{\pi}{2},0)$ the calculation is again the same as for the $1^{st}$ column
except that $\xi_n^{(x)}$ has to be replaced by $\xi_n^{(y)}$ and $\eta_n^{(x)}$ by $\eta_n^{(y)}$.
As mentioned before, the $3^{rd}$ column of $R(\frac{\pi}{2},0)$ is the vector product of the $1^{st}$ and the $2^{nd}$ ones.
Since we only need a particular matrix element of the half period monodromy, namely $R(\pi,0)_{3,3}$,
it suffices to use the following equation resulting from (\ref{R14}):
\begin{equation}\label{FS8}
 R(\pi,0)_{3,3}=R({\textstyle\frac{\pi}{2}},0)_{1,3}^2-R({\textstyle\frac{\pi}{2}},0)_{2,3}^2+R({\textstyle\frac{\pi}{2}},0)_{3,3}^2
 =1-2\,R({\textstyle\frac{\pi}{2}},0)_{2,3}^2
 \;,
\end{equation}
and hence
\begin{equation}\label{FS9}
  r=\left| R({\textstyle\frac{\pi}{2}},0)_{2,3}\right|
  =\left|
   R({\textstyle\frac{\pi}{2}},0)_{3,1}\, R({\textstyle\frac{\pi}{2}},0)_{1,2}- R({\textstyle\frac{\pi}{2}},0)_{1,1}\, R({\textstyle\frac{\pi}{2}},0)_{3,2}
  \right|
  \;,
\end{equation}
where the entries from the first and second column of $ R({\textstyle\frac{\pi}{2}},0)$ have been calculated above.
This completes the determination of the auxiliary parameter $r$ and the quasienergy via (\ref{Q8}).

\begin{figure}[h]
  \centering
    \includegraphics[width=0.75\linewidth]{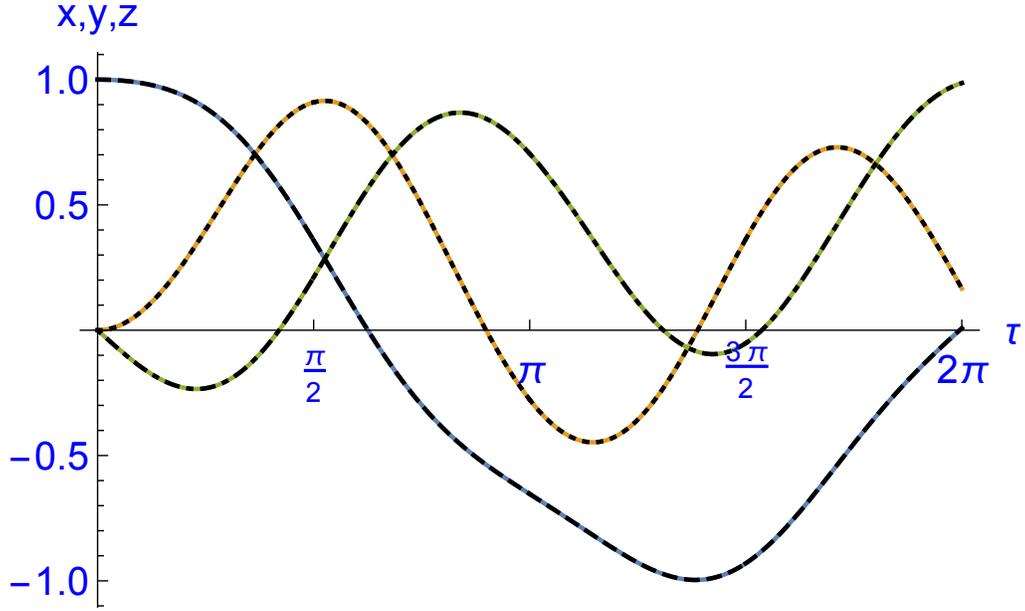}
  \caption[T1]
  {The three components of the classical spin vector as functions of dimensionless time $\tau$ over one period
  according to the equation of motion (\ref{D1}).
  We have chosen the  parameters $f=1,\,g=1/2$ and $\nu=1$ and the initial conditions $x(0)=1,\;y(0)=z(0)=0$.
  The solid curves are numerical results; the dashed curve represents $x(\tau)$ as calculated analytically,
  likewise $y(\tau)$ (dotted curve) and $z(\tau)$ (dotted-dashed curve).
   }
  \label{FIGPNA}
\end{figure}
\begin{figure}[h]
  \centering
    \includegraphics[width=0.75\linewidth]{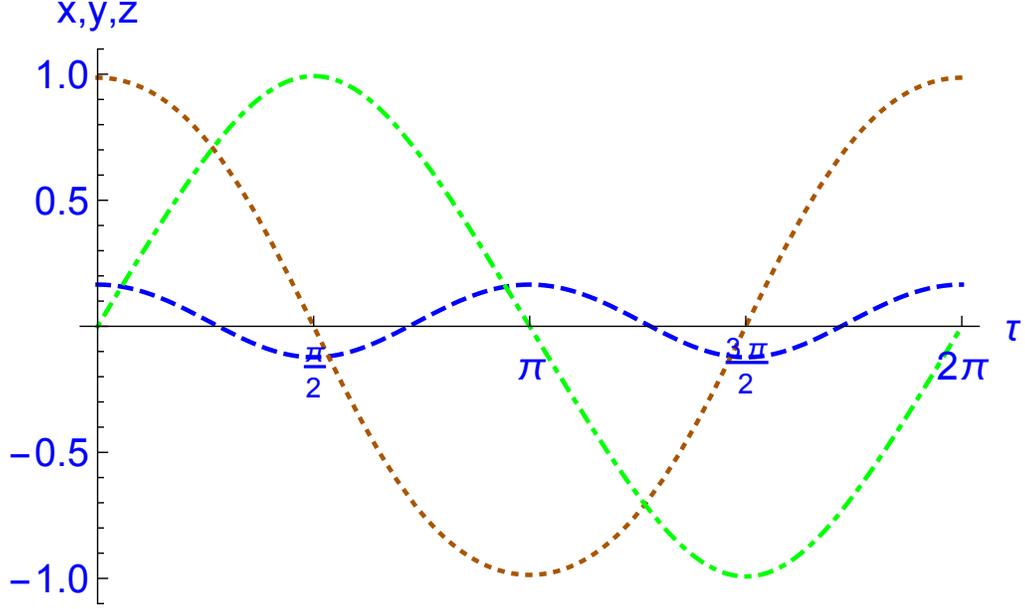}
  \caption[T1]
  {The three components of the classical spin vector as periodic functions of dimensionless time $\tau$ over one period
  according to the equation of motion (\ref{D1}).
  We have chosen the  parameters $f=1,\,g=1/2$ and $\nu=1$ and the initial conditions $x(0)=\cos\alpha,\;y(0)=\sin\alpha$ and $z(0)=0$.
  The dashed curve represents $x(\tau)$ as calculated numerically and  analytically,
  likewise $y(\tau)$ (dotted curve) and $z(\tau)$ (dotted-dashed curve).
   }
  \label{FIGPNB}
\end{figure}

We have checked the results (\ref{FS4}) and (\ref{FS9}) by comparison with a numerical solution of the $s=\frac{1}{2}$
Schr\"odinger equation (\ref{SE5}) for the choice of the parameters $\nu=1,\;f=1,$ and $g=1/2$.  For this case both methods come to the
same conclusion
\begin{equation}\label{FS10}
\alpha=1.40464\ldots,\quad r=0.387328\ldots,\quad \mbox{and hence}\quad   \epsilon^{(qu)}=0.126602\ldots
\;.
\end{equation}

%%%%%%%%%%%%%%%%%%%%%%%%%%%%%%%%%%%%%%%%%%%%%%%%%%%%%%%%%%%%%%%%%%%%%%%%%%%%%%%%%%%%%%%%%%%%%%%%%%%%%%%%%%%%%%%%%%%%%%%%%%%%%%%
\section{Time evolution: An example}\label{sec:T}
%%%%%%%%%%%%%%%%%%%%%%%%%%%%%%%%%%%%%%%%%%%%%%%%%%%%%%%%%%%%%%%%%%%%%%%%%%%%%%%%%%%%%%%%%%%%%%%%%%%%%%%%%%%%%%%%%%%%%%%%%%%%%%%
%%%%%%%%%%%%%%%%%%%%%%%%%%%%%%%%%%%%%%%%%%%%%%%%%%%%%%%%%%%%%%%%%%%%%%%%%%%%%%%%%%%%%%%%%%%%%%%%%%%%%%%%%%%%%%%%%%%%%%%%%%%%

As an example we consider the time evolution over one period $\tau\in[0,2\pi]$ according to (\ref{D1}).
We choose the values of the parameters $f=1,\,g=1/2$ and $\nu=1$ and analytically
calculate three mutually orthogonal solutions ${\mathbf S}^{(i)}(\tau),\;i=1,2,3$ for $\tau\in[0,\frac{\pi}{2}]$
by evaluating the corresponding power series solutions with ten terms. For the remaining three quarter periods
we adopt the reduction equations (\ref{R3}) and (\ref{R8}) for ${\mathbf S}^{(1)}$, where
$R(\pi,0)$ can be expressed through  $R({\textstyle\frac{\pi}{2}},0)$ via (\ref{R14}).
We observe a satisfactory agreement with the numerical solution of (\ref{D1}), see Figure \ref{FIGPNA}.

The alternative choice of the initial conditions as $x(0)=\cos\alpha$ and $y(0)=\sin\alpha$, whereas $z(0)=0$ remains unchanged,
leads to $2\pi$-periodic solutions, see Figure \ref{FIGPNB}.
This calculation uses the value of the auxiliary parameter $\alpha$
that has been determined according to (\ref{FS4}).

The first few terms of the corresponding Fourier series read:
\begin{eqnarray}
\label{T2a}
  x(\tau) &=& 0.0240019\, +0.144012 \cos (2 t)-0.00263811 \cos (4 t)+\ldots\\
  \label{T2b}
  y(\tau) &=& 1.01784 \cos (t)-0.0319147 \cos (3 t)+0.000303923 \cos (5 t)+\ldots \\
  \label{T2c}
  z(\tau) &=& 0.969835 \sin (t)-0.0224197 \sin (3 t)+0.000192725 \sin (5 t)+\ldots
  \;.
\end{eqnarray}

%%%%%%%%%%%%%%%%%%%%%%%%%%%%%%%%%%%%%%%%%%%%%%%%%%%%%%%%%%%%%%%%%%%%%%%%%%%%%%%%%%%%%%%%%%%%%%%%%%%%%%%%%%%%%%%%%%%%%%%%%%%%%%%
\section{Vanishing of the Quasienergy}\label{sec:QIII}
%%%%%%%%%%%%%%%%%%%%%%%%%%%%%%%%%%%%%%%%%%%%%%%%%%%%%%%%%%%%%%%%%%%%%%%%%%%%%%%%%%%%%%%%%%%%%%%%%%%%%%%%%%%%%%%%%%%%%%%%%%%%%%%

\begin{figure}[h]
  \centering
  \includegraphics[width=0.75\linewidth]{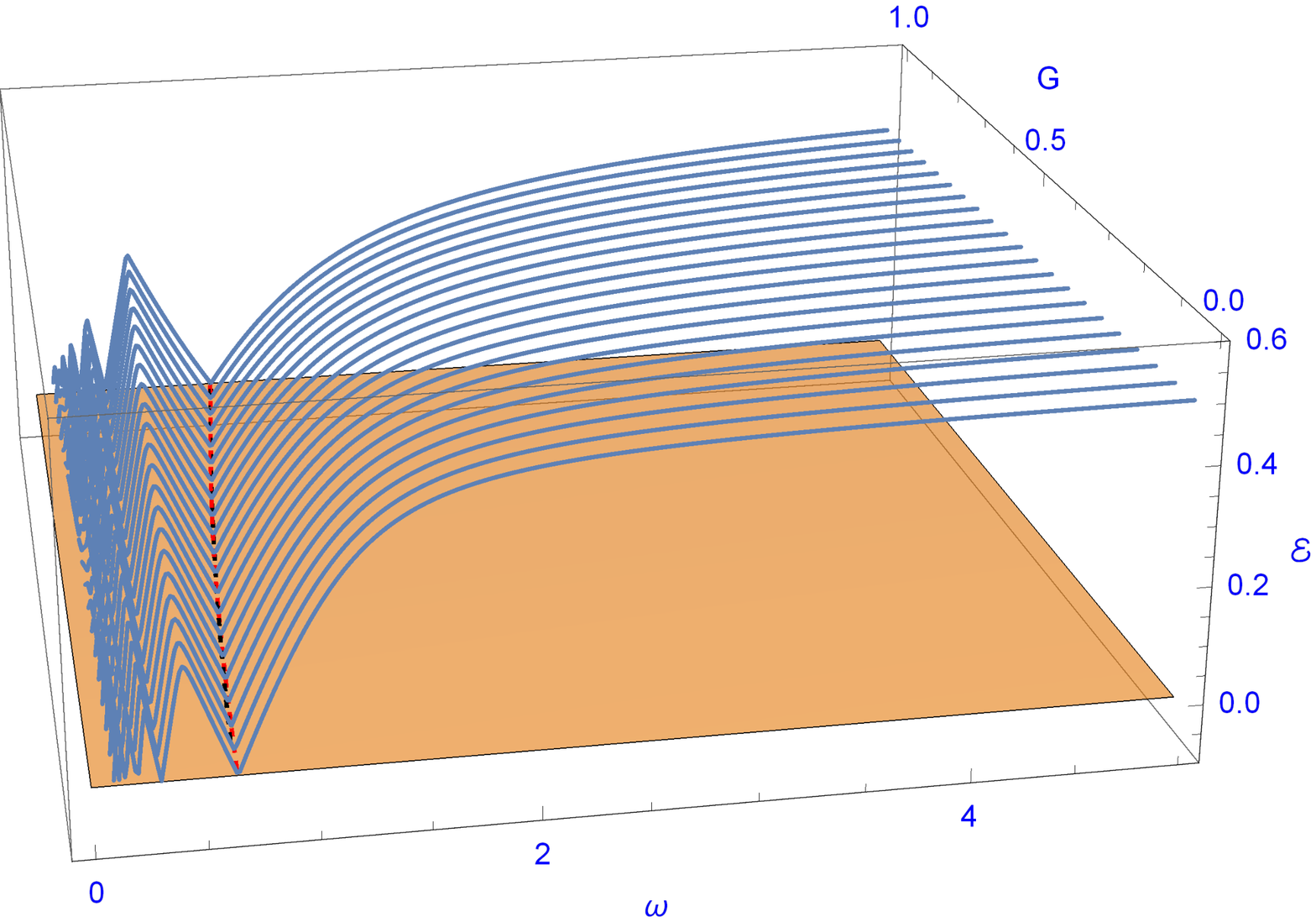}
  \caption[T1]
  {The branch of the quasienergy ${\mathcal E}(\omega_0,F,G,\omega)$ satisfying (\ref{QIIIbranch}) as a function of $\omega$
  and $G$ for fixed values of $\omega_0=1$ and $F=1$. $G$ varies from $G=0$ (linear polarization) to $G=F=1$ (circular polarization).
  Along the red, dashed curve the quasienergy vanishes. An analytical approximation of this curve according to (\ref{QIII3}) is shown
  as a black, dashed curve.
   }
  \label{FIGSFP}
\end{figure}

\begin{figure}[h]
  \centering
  \includegraphics[width=0.75\linewidth]{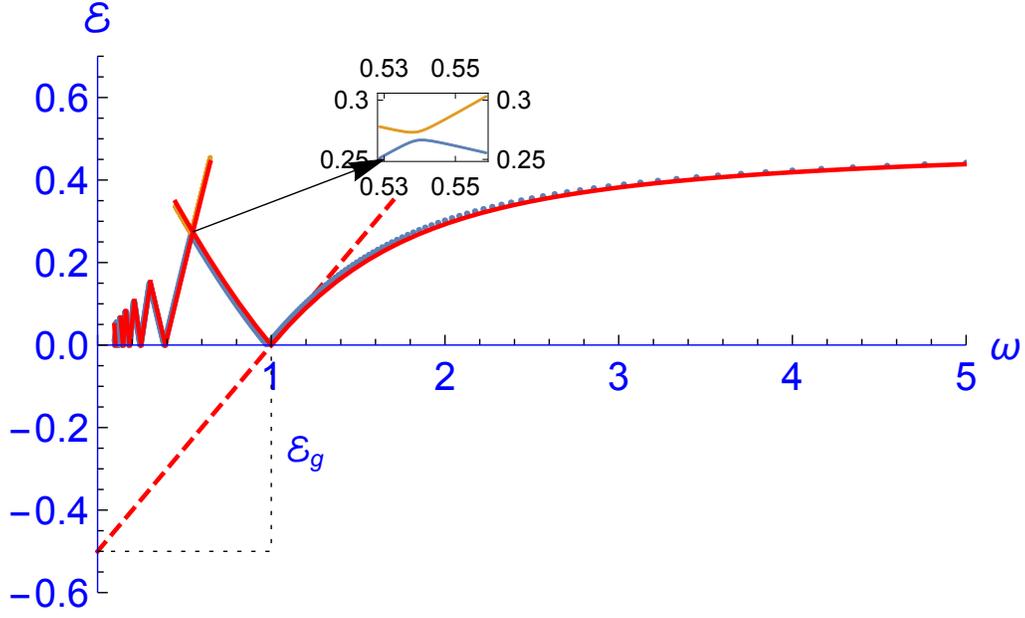}
  \caption[T1]
  {The branch of the quasienergy ${\mathcal E}(\omega_0,F,G,\omega)$ satisfying (\ref{QIIIbranch}) as a function of $\omega$
  for fixed $\omega_0=1,\,F=1$ and two values $G=1$ (red curves) and $G=0.95$ (blue and orange curves).
  In the circular case ($G=1$) we observe level crossing whereas in the case with small eccentricity ($G=0.95$) this crossing
  is avoided as demonstrated by the inset. The dashed red line is the tangent of ${\mathcal E}(1,1,1,\omega)$ at $\omega=1$.
  According to (\ref{QIIIslope}) it has the slope ${\mathcal E}_g=\frac{1}{2}$.
   }
  \label{FIGS1}
\end{figure}

\begin{figure}[h]
  \centering
  \includegraphics[width=0.75\linewidth]{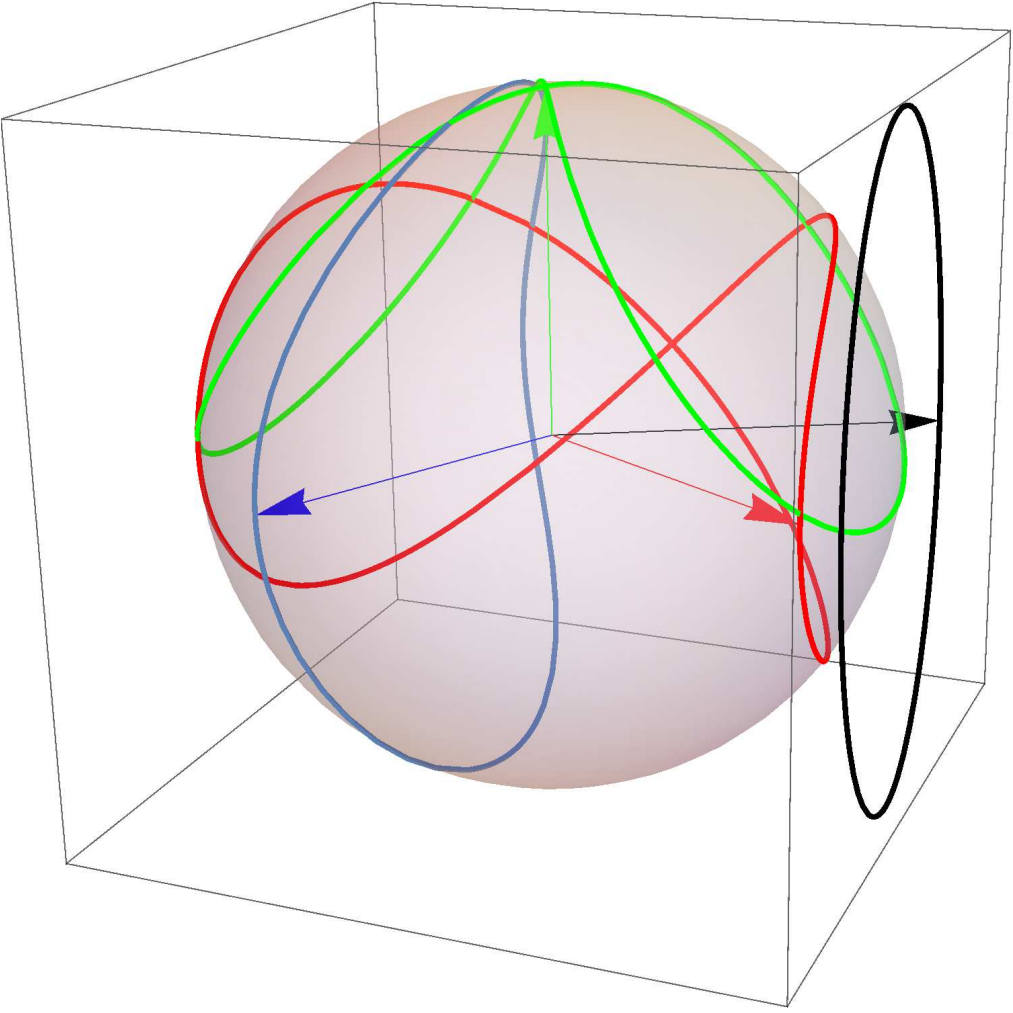}
  \caption[T1]
  {Three periodic solutions ${\mathbf S}^{(i)}(\tau),\;i=1,2,3$ (blue, red and green curves) of (\ref{D1}) for the parameters $\omega_0=F=1,\;G=\frac{1}{2}$ and
  $\omega=\omega_1$ such that the quasienergy vanishes. The magnetic field vector (black arrow) moves on the black ellipse;
  the initial vectors ${\mathbf S}^{(i)}(0),\;i=1,2,3$ are shown as colored arrows. The time average of the energy vanishes for
  ${\mathbf S}^{(2)}$ and ${\mathbf S}^{(3)}$.
   }
  \label{FIGSP12}
\end{figure}

\begin{figure}[h]
  \centering
  \includegraphics[width=0.75\linewidth]{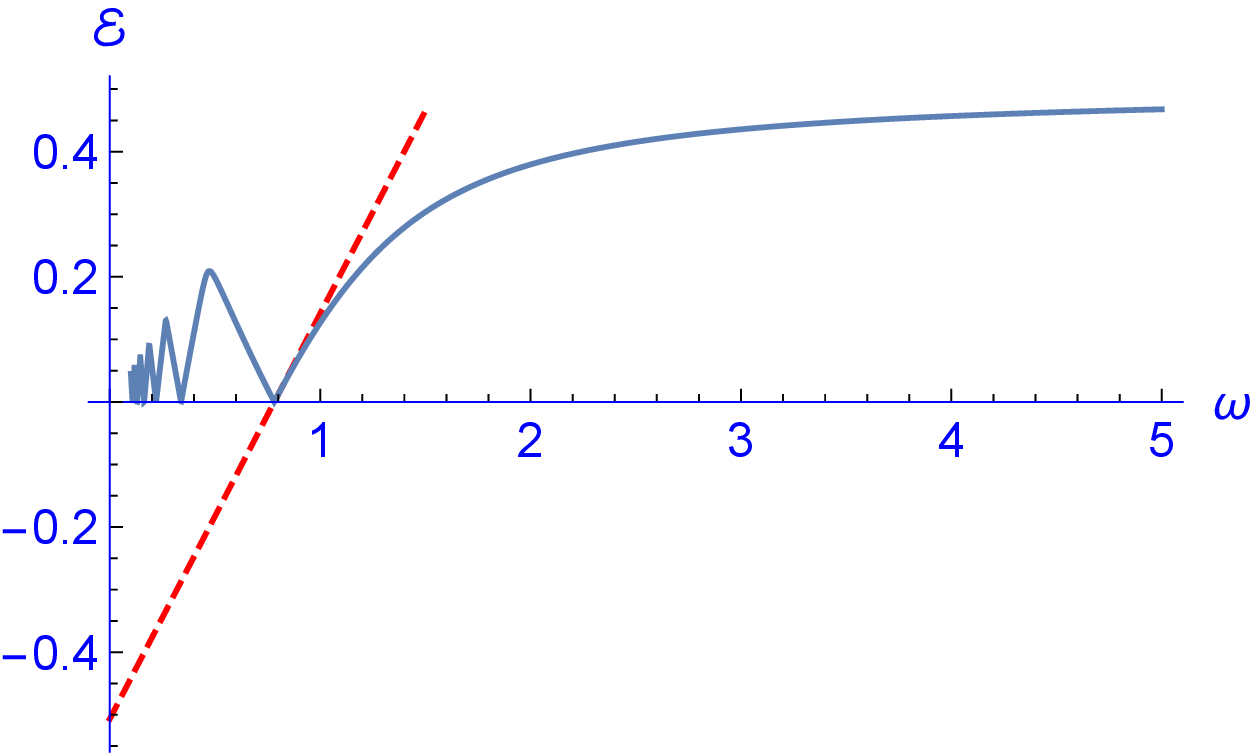}
  \caption[T1]
  {The  branch of the quasienergy ${\mathcal E}(\omega_0,F,G,\omega)$ satisfying (\ref{QIIIbranch}) as a function of $\omega$
  for fixed values of $\omega_0=F=1$ and $G=\frac{1}{2}$. The function has its largest zero at $\omega_1=0.781665$
  where the tangent (dashed red line) has a slope of $0.64787$, see the text after Eq.~(\ref{QIIIedb}).
   }
  \label{FIGS2}
\end{figure}

We will discuss the quasienergy in physical units
\begin{equation}\label{QIII1}
 {\mathcal E}(\omega_0,F,G,\omega)\equiv\hbar\,\omega\, \epsilon^{(qu)}\left(\frac{\omega_0}{\omega},\frac{F}{\omega},\frac{G}{\omega}\right)
 = \hbar\,\omega \,\epsilon^{(qu)}\left(\nu,f,g\right)
 \;,
\end{equation}
where usually $\hbar$ is set to $1$. Analogously to the ambiguity of $\epsilon^{(qu)}$ also ${\mathcal E}$ will be only defined
up to integer multiples of $\hbar\omega$.
A typical plot of the functions $\omega\mapsto  {\mathcal E}(\omega_0,F,G,\omega)$ for the values $\omega_0=F=1$ and $G$ varying
from $G=0$ (linear polarization) to $G=F=1$ (circular polarization) is shown in Figure \ref{FIGSFP}, where the branch and the sign
of the quasienergy are chosen according to
\begin{equation}\label{QIIIbranch}
  0\le \frac{1}{\hbar\omega} {\mathcal E}(1,1,G,\omega)\le \frac{1}{2}
  \;.
\end{equation}
We notice that these curves qualitatively all look the same. First we note that the family of curves shows the same asymptotic behaviour of
${\mathcal E}(\omega_0,F,G,\omega)$ for $\omega\rightarrow\infty$. In the case of circular polarization we have
${\mathcal E}(\omega_0,F,G,\omega)\rightarrow \frac{\omega_0}{2}$ for $G=F$
as well as  ${\mathcal E}(\omega_0,F,G,\omega)\rightarrow\frac{\omega_0}{2}$ for linear polarization, see Eq.~(269) in \cite{S18}.
Further, the quasienergy functions have an infinite number of zeros with a non-vanishing slope, the largest being slightly below $\omega=1$.\\

To better understand this behavior in detail we re-visit the RPC.
It is well-known that in the special case of circular polarization the quasienergy can be analytically determined in a rather simple form.
In the context of the present discussion we note that the fundamental matrix solution of (\ref{D3}) with initial condition  (\ref{D4})
assumes the form $R(\tau,0)=\left( R_1, R_2, R_3\right)$ with the three columns reading
\begin{eqnarray}
\label{QIIIColumn1}
  R_1 &=& \left(\begin{array}{c}
                  f^2 \cos (\tau  \Omega )+(\nu -1)^2 \\
 f \left(2 (\nu -1) \cos (\tau ) \sin ^2\left(\frac{\tau  \Omega }{2}\right)+\Omega
    \sin (\tau ) \sin (\tau  \Omega )\right)
   \\
 f \left(2 (\nu -1) \sin (\tau ) \sin ^2\left(\frac{\tau  \Omega }{2}\right)-\Omega
    \cos (\tau ) \sin (\tau  \Omega )\right)
                \end{array} \right) \;,\\
\label{QIIIColumn2}
  R_2&=& \left(\begin{array}{c}
                   -f (\nu -1) (\cos (\tau  \Omega )-1)  \\
  \cos (\tau ) \left(f^2+(\nu -1)^2
   \cos (\tau  \Omega )\right)-(\nu -1) \Omega  \sin (\tau ) \sin (\tau  \Omega )
   \\
\sin (\tau ) \left(f^2+(\nu -1)^2
   \cos (\tau  \Omega )\right)+(\nu -1) \Omega  \cos (\tau ) \sin (\tau  \Omega )
                \end{array} \right) \;,    \\
\label{QIIIColumn3}
 R_3&=& \left(\begin{array}{c}
                 f \Omega  \sin (\tau  \Omega ) \\
 -\Omega  ((\nu -1) \cos (\tau ) \sin (\tau  \Omega )+\Omega  \sin (\tau ) \cos
   (\tau  \Omega )) \\
 \Omega  (\Omega  \cos (\tau ) \cos (\tau  \Omega )-(\nu -1) \sin (\tau ) \sin
   (\tau  \Omega )) \\
                \end{array} \right) \;,
\end{eqnarray}
where we have used the abbreviation
\begin{equation}\label{QIIIOmega}
  \Omega\equiv \sqrt{f^2+(1-\nu)^2}
  \;,
\end{equation}
known as the ``Rabi frequency".
The corresponding monodromy matrix $R(2\pi,0)$ reads:
\begin{equation}\label{QIIImono}
 R(2\pi,0) =\left(
\begin{array}{ccc}
 \frac{f^2 \cos (2 \pi  \Omega )+(\nu -1)^2}{\Omega ^2} & \frac{2 f (\nu -1) \sin
   ^2(\pi  \Omega )}{\Omega ^2} & \frac{f \sin (2 \pi  \Omega )}{\Omega } \\
 \frac{2 f (\nu -1) \sin ^2(\pi  \Omega )}{\Omega ^2} & \frac{f^2+(\nu -1)^2 \cos (2
   \pi  \Omega )}{\Omega ^2} & -\frac{(\nu -1) \sin (2 \pi  \Omega )}{\Omega } \\
 -\frac{f \sin (2 \pi  \Omega )}{\Omega } & \frac{(\nu -1) \sin (2 \pi  \Omega
   )}{\Omega } & \cos (2 \pi  \Omega ) \\
\end{array}
\right).
%\normalsize
\end{equation}
Its trace is evaluated as $\mbox{Tr}\left( R(2\pi,0)\right)=1+2\,\cos(2\pi\,\Omega)$
and yields the eigenvalues $(1,\exp(\pm 2\,\pi\,{\sf i} \,\Omega))$,
 corresponding to a classical quasienergy
\begin{equation}\label{QIIIepscl}
 \epsilon^{(cl)}=\Omega
 \;.
\end{equation}
On the other hand, we may apply (\ref{D8}) to the
periodic solution
\begin{equation}\label{QIIIperiodic}
 {\mathbf S}(\tau)=\frac{1}{\Omega}\left(
 \begin{array}{c}
   \nu -1 \\
   f\,\cos \tau \\
   f\,\sin\tau
 \end{array}
 \right)
\end{equation}
with the well-known result
\begin{equation}\label{QIIIepsqu}
 \epsilon^{(qu)}=\frac{1\pm \Omega}{2}
 \;,
\end{equation}
that is compatible with (\ref{QIIIepscl}) and (\ref{D7}).

For $G=F=\omega_0=1$ the quasienergy curve has a zero at $\omega=1$, i.~e., ${\mathcal E}(1,1,1,1)=0$.
For slightly lower values of $G$ this zero shifts to lower values of $\omega$, see Figures \ref{FIGSFP} and \ref{FIGS1}.
We will denote by $G={\mathcal E}_0(\omega; F,\omega_0)$ the position of the largest zero of the quasienergy.

The vanishing of the quasienergy is in so far interesting as it means that {\it all} solutions of (\ref{D1})
will be $2\pi$-periodic, not only the special one with the initial condition
${\mathbf S}(0)=(\cos\alpha,\sin\alpha,0)^\top$ according to (\ref{Q9}).
Moreover,  ${\mathcal E}(\omega_0,F,G,\omega)=0$ means degeneracy of the Floquet states for the two level system,
which may produce some $2^{nd}$ order phase transition in the parameter space, see  \cite{S20a}.

The vanishing of the quasienergy implies that the linear term $z_0\,\tau$ in (\ref{FS2d}) must vanish and hence
\begin{equation}\label{QIII2}
  0=z_0\stackrel{(\ref{FS2e})}{=}\nu\,y_0-\frac{g}{2}\,x_1
  \;.
\end{equation}
In order to check the consistency we will evaluate the condition (\ref{QIII2}) by using a truncation of the power series
solutions (\ref{H8e})  and (\ref{H12e}) to the first ten terms. This yields the exact first five terms of $G={\mathcal E}_0(\omega;1,1)$ expanded
into a power series in terms of $\omega-\omega_0=\omega-1$:
\begin{equation}\label{QIII3}
 {\mathcal E}_0(\omega;1,1)=1+2 (\omega-1)-\frac{5}{6} (\omega-1)^2+\frac{49}{36} (\omega-1)^3-\frac{577}{240} (\omega-1)^4
 +\frac{58357}{12960}(\omega-1)^5+O\left((\omega -1)^6\right)
 \;.
\end{equation}
The result is shown in Figure \ref{FIGSFP} as a black dashed curve and fits to the numerically
determined red dashed curve of vanishing quasienergy in the domain $0.7<\omega<1$.

Further we note that according to \cite{S18} the quasienergy ${\mathcal E}$ can be split into a geometrical part
${\mathcal E}_g$ and a dynamical part ${\mathcal E}_d$ such that ${\mathcal E}={\mathcal E}_g+{\mathcal E}_d$
and the slope relation
\begin{equation}\label{QIIIslope}
 \frac{\partial {\mathcal E}}{\partial \omega }=\frac{{\mathcal E}_g}{\omega}
\end{equation}
holds, see Eq.~(151) in \cite{S18}. Recall that ${\mathcal E}_d$ is the time average of the energy, i.~e,
${\mathcal E}_d=\frac{1}{2}\overline{{\mathbf h}(t)\cdot{\mathbf S}(t)}$ and
${\mathcal E}_g=\frac{\omega}{4\pi}\,\left|{\mathcal A}\right|$, where $\left|{\mathcal A}\right|$
denotes the signed area of the Bloch sphere swept by ${\mathbf S}(t)$ over one period.
In our case this implies that for vanishing quasienergy and hence $G={\mathcal E}_0(\omega)$ we have
${\mathcal E}_g+{\mathcal E}_d=0$ and  the slope of the curve $\omega\mapsto {\mathcal E}(\omega_0,F,G,\omega)$
equals $\left| {\mathcal E}_g\right|=\left| {\mathcal E}_d\right|$. We have illustrated this relation
for the special case of circular polarization in Figure \ref{FIGS1} by drawing the tangent (dashed red line)
with the slope $\frac{1}{2}$. This corresponds to a periodic solution of (\ref{D1}) tracing a great circle
on the Bloch sphere with solid angle $\left|{\mathcal A}\right|=2\,\pi$.

In general the quasienergy (\ref{QIIIepsqu}) of the RPC has its first zero
at $\omega =\omega_1\equiv\frac{F^2+\omega _0^2}{2 \omega _0}$. Hence the series (\ref{QIII3}) will assume
its general form as $G={\mathcal E}_0(\omega;F,\omega_0)=F+\sum_{n=1}^\infty g_n\,(\omega-\omega_1)^n$.
It begins with
\begin{equation}\label{QIII4}
 G= {\mathcal E}_0(\omega)= F-\frac{6 \left(-60 F^4+50 F^4 \omega _0-100 F^2 \omega _0^2+71 F^2 \omega _0^3-25\omega _0^4\right)
   \left(\omega -\frac{F^2+\omega _0^2}{2 \omega _0}\right)}{F\left(30 F^4+20 F^2 \omega _0+67 F^2 \omega _0^2+10 \omega _0^3
   +65 \omega_0^4\right)}+\ldots\;,
\end{equation}
but the next terms are too intricate to be shown here.

We will consider the case of vanishing quasienergy along the curve $G={\mathcal E}_0(\omega)$ in more details.
As already mentioned, in this case all solutions of (\ref{D1}) will be $2\pi$-periodic and hence the local
Fourier series representations (\ref{FSfourx}), (\ref{FSfoury}) and (\ref{FS2d}) can be extended to all times $\tau$.
It turns out that the slope relation (\ref{QIIIslope}) cannot be satisfied by {\it all} periodic solutions of (\ref{D1}) but
only by a particular one that is the limit of the (up to a sign) unique periodic solutions for non-vanishing quasienergy.
Instead of again using Eq.~(\ref{FS4}) to determine this limit we will proceed in a different way.

Recall that for a general periodic, not necessarily normalized solution ${\mathbf S}(\tau)=(x(\tau),y(\tau),z(\tau))^\top$
of (\ref{D1}) of the form (\ref{FSfourx}), (\ref{FSfoury}) and (\ref{FSfourz})
the functions $x(\tau)$ and $y(\tau$) are represented by  $\cos$-series whereas $z(\tau)$ will be a $\sin$-series.
Consider the decomposition of $x(\tau)=x_e(\tau)+x_o(\tau)$ into an even $\cos$-series and an odd one and analogously for
$y(\tau)=y_e(\tau)+y_o(\tau)$ and $z(\tau)=z_e(\tau)+z_o(\tau)$. Consequently, the time derivative $\dot{x}$ can be
uniquely split into an even $\sin$-series and an odd one:
\begin{equation}\label{QIIIsplit}
  \dot{x}=\dot{x_e}+\dot{x_o}= \left( g\,\cos(\tau)\,z_o-f\,\sin(\tau)\,y_o\right)+\left( g\,\cos(\tau)\,z_e-f\,\sin(\tau)\,y_e\right)
  \;.
\end{equation}
Analogous decompositions for $\dot{y}$ and $\dot{z}$ lead to a decomposition of ${\mathbf S}(\tau)$ into
two separate solutions
\begin{equation}\label{QIIIdec}
 {\mathbf S}(\tau)= {\mathbf S}^{(1)}(\tau)+ {\mathbf S}^{(2)}(\tau)\equiv
 \left(\begin{array}{c}
         x_e(\tau) \\
         y_o(\tau) \\
         z_o(\tau)
       \end{array}
 \right)+
 \left(\begin{array}{c}
         x_o\tau) \\
         y_e(\tau) \\
         z_e(\tau)
       \end{array}
 \right)
 \;.
\end{equation}
It is clear that ${\mathbf S}^{(1)}(\tau)$ equals the limit of periodic solutions for non-vanishing
quasienergies since these periodic solutions have the same even/odd character as  ${\mathbf S}^{(1)}(\tau)$,
see Proposition \ref{prop5}. Moreover, the two solutions (\ref{QIIIdec}) are orthogonal for all $\tau$:
Their scalar product is constant in time, on the other hand an odd $\cos$-series and has thus a vanishing time average.
Further, it follows that ${\mathbf S}^{(2)}(\tau)$ belongs to ${\mathcal E}_d=0$ since
${\mathbf h}\cdot {\mathbf S}^{(2)}$ will be an odd $\cos$-series and thus has a vanishing time average.
In contrast, ${\mathbf h}\cdot {\mathbf S}^{(1)}$ will be an even $\cos$-series which is compatible
with $\left|{\mathcal E}_d\right|>0$ and a positive slope of the quasienergy curves at $G={\mathcal E}_0(\omega)$,
see Figure \ref{FIGSFP}.

For the sake of completeness we note that the third solution ${\mathbf S}^{(3)}= {\mathbf S}^{(1)}\times {\mathbf S}^{(2)}$
will be of the following type: $x^{(3)}(\tau)$ is an odd $\sin$-series, $y^{(3)}(\tau)$ is an even $\sin$-series, and
$z^{(3)}(\tau)$ is an even $\cos$-series. Hence also for this solution the time average of ${\mathbf h}\cdot {\mathbf S}^{(3)}$
will be an odd $\sin$-series and hence ${\mathcal E}$ vanishes. An example is shown in Figure \ref{FIGSP12}.

In the case of the periodic solutions ${\mathbf S}^{(1)}(\tau)$ or ${\mathbf S}^{(2)}(\tau)$
the time average of ${\mathbf h}\cdot {\mathbf S}$ can be expressed
in terms of the first Fourier coefficients:
\begin{equation}\label{QIIIav}
 \epsilon_d^{(qu)}=\frac{1}{2}\overline{ {\mathbf h}\cdot {\mathbf S}}=\frac{1}{2}
 \left( \nu\,x_0+\frac{g}{2}y_1+\frac{f}{2}z_1\right)
 \;,
\end{equation}
where we have again passed to the dimensionless quasienergy and the Fourier coefficients are given in
(\ref{FS2a}),  (\ref{FS2b}) and (\ref{FS2d}). The suitable initial conditions $x(0)=\cos\beta$ and $y(0)=\sin\beta$
can be derived from the result  $\epsilon_d^{(qu)}=0$ analogously to (\ref{FS4}):
\begin{equation}\label{QIIIbeta}
\beta=-\arctan\frac{ \nu\,x_0^{(x)}+\frac{g}{2}y_1^{(x)}+\frac{f}{2}z_1^{(x)}}{ \nu\,x_0^{(y)}+\frac{g}{2}y_1^{(y)}+\frac{f}{2}z_1^{(y)}}
\;.
\end{equation}
Here the superscript $(x)$ or $(y)$ refers to the dependence of the Fourier coefficients, via $\xi_n$ and $\eta_n$, on the
initial conditions $x(0)$ and $y(0)$. Finally, the initial conditions  $x(0)=\cos\alpha$ and $y(0)=\sin\alpha$ for the first
solution $(\mathbf S)^{(1)}(\tau)$  are given by
\begin{equation}\label{QIIIalpha}
  \alpha=\beta\pm\frac{\pi}{2}
  \;,
\end{equation}
using the orthogonality of ${\mathbf S}^{(1)}(\tau)$ and ${\mathbf S}^{(2)}(\tau)$. After some calculations it follows that
the dynamical part of the quasienergy of the first solution ${\mathbf S}^{(1)}(\tau)$ assumes the value
\begin{eqnarray}\label{QIIIeda}
 \epsilon_d&=&\frac{1}{2}\overline{ {\mathbf h}\cdot {\mathbf S}^{(1)}}=\frac{1}{2}
 \left(
 \left(\nu\,x_0^{(x)}+\frac{g}{2}y_1^{(x)}+\frac{f}{2}z_1^{(x)}
 \right)\cos\alpha+
 \left(\nu\,x_0^{(y)}+\frac{g}{2}y_1^{(y)}+\frac{f}{2}z_1^{(y)}
 \right)\sin\alpha
 \right)\\
 \label{QIIIedb}
 &=&
 \frac{1}{2}
 \sqrt{
 \left(\nu\,x_0^{(x)}+\frac{g}{2}y_1^{(x)}+\frac{f}{2}z_1^{(x)}
 \right)^2+
 \left(\nu\,x_0^{(y)}+\frac{g}{2}y_1^{(y)}+\frac{f}{2}z_1^{(y)}
 \right)^2
 }\;.
\end{eqnarray}

As an example we consider the parameters $\omega_0=F=1$ and $G=\frac{1}{2}$. The quasienergy curve
$\omega\mapsto {\mathcal E}(1,1,\frac{1}{2},\omega)$ has its largest zero at $\omega_1=0.781665$.
This value has been determined numerically; the analytical approximation (\ref{QIII3}) yields $\omega_1=0.781023$.
At this point the two solutions ${\mathbf S}^{(1)}$ and ${\mathbf S}^{(2)}$ are obtained with initial values
$x(0)=\cos\alpha,\;y(0)=\sin\alpha$, and $x(0)=\cos\beta,\;y(0)=\sin\beta$, resp.~, where $\beta=-0.489254$ has been
calculated according to (\ref{QIIIbeta}) and $\alpha=\beta+\frac{\pi}{2}$, see Figure \ref{FIGSP12}.
The slope of the tangent of the quasienergy curve at $\omega=\omega_1$ has been determined via (\ref{QIIIedb}) and assumes the value
$\epsilon_d=0.64787$, see Figure \ref{FIGS2}.

%%%%%%%%%%%%%%%%%%%%%%%%%%%%%%%%%%%%%%%%%%%%%%%%%%%%%%%%%%%%%%%%%%%%%%%%%%%%%%%%%%%%%%%%%%%%%%%%%%%%%%%%%%%%%%%%%%%%%%%%%%%%%%
\section{Resonances}\label{sec:RES}
%%%%%%%%%%%%%%%%%%%%%%%%%%%%%%%%%%%%%%%%%%%%%%%%%%%%%%%%%%%%%%%%%%%%%%%%%%%%%%%%%%%%%%%%%%%%%%%%%%%%%%%%%%%%%%%%%%%%%%%%%%%%%%%%%%%%%%

\begin{figure}[h]
  \centering
  \includegraphics[width=0.75\linewidth]{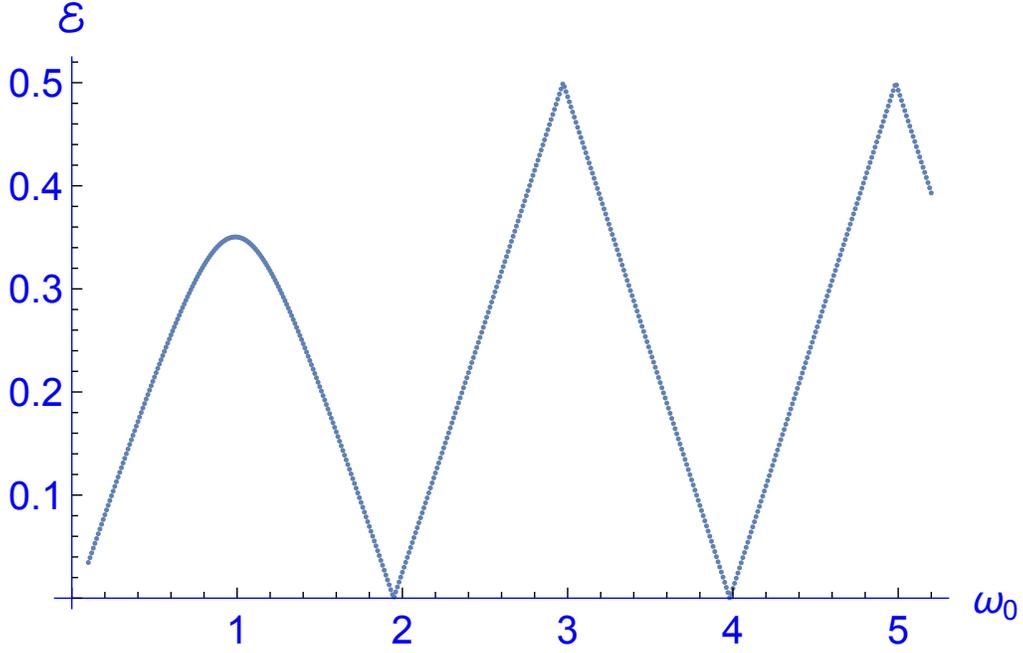}
  \caption[T1]
  {The quasienergy ${\mathcal E}(\omega_0,F,G,\omega)$ as a function of $\omega_0$ for fixed values of $\omega=1$, $F=0.5$ and $G=0.1$
  (solid curves) calculated by numerical solutions of the Schr\"odinger equation (\ref{SE5}).
  One observes maxima of the quasienergy at $\omega_0\approx 1,3,5,\ldots$. The dotted curves are various branches of the analytical
  form (\ref{RESC1}) of the quasienergy for the case of circular polarization, i.~e., $G=0$ and $\omega=1$, $F=0.5$.
   }
  \label{FIGRES1}
\end{figure}

The function $\omega_0\mapsto {\mathcal E}(\omega_0,F,G,\omega)$, restricted to the domain (\ref{QIIIbranch}), has an infinite number of maxima,
see Figure \ref{FIGRES1}. These satisfy the condition
\begin{equation}\label{RES0}
 0=\frac{\partial }{\partial \omega_0} {\mathcal E}(\omega_0,F,G,\omega)
 \;.
\end{equation}
defining an infinite number of hypersurfaces in the parameter space ${\mathcal P}$ with points $(\omega_0,F,G,\omega)\in{\mathcal P}$.
Solving (\ref{RES0}) for $\omega$ gives the so-called ``resonance frequencies"
\begin{equation}\label{RES01}
 \omega=\omega_{\text{res}}^{(n)}\left(\omega_0,F,G\right),\; n=1,2,3,\ldots
 \;.
\end{equation}
In the circular case a smooth representative of the quasienergy ${\mathcal E}_c$ assumes the form
\begin{equation}\label{RESC1}
 {\mathcal E}_c={\textstyle\frac{1}{2}}\left( \omega-\Omega\right)=
 {\textstyle\frac{1}{2}} \left(\omega -\sqrt{F^2+\left(\omega -\omega _0\right){}^2}\right)
 \;,
 \end{equation}
 and has a unique maximum at $\omega_0=\omega$, see Figure \ref{FIGRES1}. This conforms with the intuitive picture
 that a resonance occurs if the driving frequency $\omega$ equals the Larmor frequency $\omega_0$ of the
 energy level splitting.
 The other maxima of the quasienergy, restricted to the domain (\ref{QIIIbranch}), are represented
 by intersections of suitable branches of the quasienergy of the form $\pm {\mathcal E}_c+n\,\omega,\; n\in{\mathbbm Z}$.
 For example, the next maximimum at  $\omega_0\approx 3\,\omega$ is obtained by the intersection
 of $-{\mathcal E}_c$ and ${\mathcal E}_c+\omega$ at
 \begin{equation}\label{RESC2}
  \omega_0=\omega+\sqrt{4\,\omega^2-F^2}= 3\,\omega-\frac{F^2}{4\,\omega}+O(F^4)
  \;.
 \end{equation}
 Note that an arbitrarily small admixture of eccentricity to the polarization leads to an avoided level crossing and
 a smooth maximum close to the value $\omega_0$ of the intersection, see Figure \ref{FIGRES1}.

According to \cite{S65}, the time average of the transition probability between different Floquet states assumes its maximum value $\overline{P}={\textstyle\frac{1}{2}}$ at the resonance frequencies, which justifies the denotation.
Although Shirley's derivation of the resonance condition refers to the RPL
case, see (1) in \cite{S65}, one can easily check that it also holds
in the more general RPE case.
Moreover, it has been shown \cite{S18} that for $\omega=\omega_{res}^{(n)}$
the classical periodic solution of (\ref{D1}) has a vanishing time-average into the direction of the constant component of the
magnetic field. According to our definitions this means that
\begin{equation}\label{RES1}
 x_0= x_0^{(x)}\,\cos\alpha +x_0^{(y)}\,\sin\alpha =0
  \;,
\end{equation}
where $\alpha$ is the auxiliary parameter leading to a periodic solution given by (\ref{FS4}).
Together with
\begin{equation}\label{RES2}
y_0=y_0^{(x)}\,\cos\alpha +y_0^{(y)}\,\sin\alpha=0
\;,
\end{equation}
see (\ref{FS3}), this  implies that the matrix
\begin{equation}\label{RES2a}
 \Xi\equiv\left(
 \begin{array}{cc}
   x_0^{(x)} & x_0^{(y)}\\
   y_0^{(x)} &  y_0^{(y)}
 \end{array}
 \right)
\end{equation}
has a non-vanishing null-vector and hence
\begin{equation}\label{RES3}
 \det\Xi= x_0^{(x)}\,y_0^{(y) }-x_0^{(y)}\,y_0^{(x)}=0
  \;.
\end{equation}
We use  truncated versions of (\ref{H8e}) and (\ref{H12e}) in order to derive the first terms of the
power series representations
\begin{equation}\label{RES4}
\frac{\omega_{res}^{(n)}}{\omega_0} = \sum_{m,k=0}^{\infty}\Omega^{(n)}_{m,k}\,\left(\frac{F}{\omega_0}\right)^m\,\left(\frac{G}{\omega_0}\right)^k
\;,
\end{equation}
analogously to \cite{S18}. We will show a few results. The first resonance $\omega_{\text{res}}^{(1)}$ is determined by
\begin{equation}\label{RES5}
\Omega^{(1)}=\left(
\begin{array}{cccccccccc}
 1 & 0 & \frac{1}{16} & 0 & \frac{1}{1024} & 0 & -\frac{35}{131072} & 0 &
   \frac{103}{8388608}&\ldots \\
 0 & -\frac{1}{8} & 0 & \frac{3}{256} & 0 & \frac{27}{65536} & 0 & -\frac{69}{262144}
   & 0&\ldots  \\
 \frac{1}{16} & 0 & -\frac{13}{512} & 0 & \frac{611}{131072} & 0 & \frac{433}{2097152}
   & 0 & * &\ldots \\
 0 & \frac{3}{256} & 0 & -\frac{315}{32768} & 0 & \frac{609}{262144} & 0 & * & 0 \\
 \frac{1}{1024} & 0 & \frac{611}{131072} & 0 & -\frac{19115}{4194304} & 0 & * & 0 & *&\ldots
   \\
 0 & \frac{27}{65536} & 0 & \frac{609}{262144} & 0 & * & 0 & * & 0 &\ldots \\
 -\frac{35}{131072} & 0 & \frac{433}{2097152} & 0 & * & 0 & * & 0 & * &\ldots \\
 0 & -\frac{69}{262144} & 0 & * & 0 & * & 0 & * & 0 &\ldots \\
 \frac{103}{8388608} & 0 & * & 0 & * & 0 & * & 0 & * &\ldots \\
 \vdots &\vdots&\vdots&\vdots&\vdots&\vdots&\vdots&\vdots&\vdots&\vdots
\end{array}
\right)
\;.
\end{equation}
We note that $\Omega^{(1)}$ is a symmetric matrix due to the symmetry of the Rabi problem
under the exchange $G\leftrightarrow F$. The matrix elements $\Omega^{(1)}_{m,k}$ vanish
for odd $m+k$. Further, it is instructive to look at the limit cases of linear or circular polarization.
For $G=0$ the first column of  $\Omega^{(1)}$ agrees with the corresponding known results
in the case of linear polarization, see Table $1$ in \cite{S18}. For $F=G$ the power series
(\ref{RES4}) coalesces into a series of a single variable $F$ with coefficients
$\tilde{\Omega}^{(1)}_M=\sum_{m=0}^{M}\Omega^{(1)}_{m,M-m},\;M=0,2,4,\ldots$.
On the other hand, the resonance frequency $\omega_{res}^{(1)}$ of the
circularly polarized case is known to be $\omega_{res}^{(1)}=\omega_0$.
Hence the anti-diagonal sums of $\Omega^{(1)}$-entries $\tilde{\Omega}^{(1)}_M$ must vanish
for $M=2,4,\ldots$. This can be confirmed for $M=0,2,\ldots,8$ in the above-shown part of $\Omega^{(1)}$, see (\ref{RES5}).

The $2^{nd}$ resonance is described by the matrix
\begin{equation}\label{RES6}
  \Omega^{(2)}=\left(
\begin{array}{ccccccccc}
 \frac{1}{3} & 0 & \frac{3}{32} & 0 & -\frac{135}{8192} & 0 & \frac{2133}{1048576} & 0
   & \ldots \\
 0 & \frac{1}{16} & 0 & -\frac{9}{2048} & 0 & -\frac{3591}{524288} & 0 & * & \ldots \\
 \frac{3}{32} & 0 & -\frac{21}{4096} & 0 & \frac{6075}{1048576} & 0 & * & 0 & \ldots \\
 0 & -\frac{9}{2048} & 0 & \frac{4095}{262144} & 0 & * & 0 & * & \ldots \\
 -\frac{135}{8192} & 0 & \frac{6075}{1048576} & 0 & * & 0 & * & 0 & \ldots \\
 0 & -\frac{3591}{524288} & 0 & * & 0 & * & 0 & * & \ldots \\
 \frac{2133}{1048576} & 0 & * & 0 & * & 0 & * & 0 & \ldots \\
 0 & * & 0 & * & 0 & * & 0 & * & \ldots \\
 \vdots &  \vdots &  \vdots &  \vdots &  \vdots &  \vdots &  \vdots &  \vdots &  \vdots \\
\end{array}
\right)\;.
\end{equation}
Here analogous remarks apply as in the case of $\Omega^{(1)}$, except that the anti-diagonal
sums of $\Omega^{(2)}$-entries $\tilde{\Omega}^{(2)}_M$  no longer vanish. They can be determined
by the following consideration. In the circular limit the $2^{nd}$ resonance is defined by the level crossing
\begin{equation}\label{RES7}
 \frac{1}{2}\left( -\omega+\Omega\right)=\frac{1}{2}\left(3\,\omega-\Omega\right)
 \;,
\end{equation}
where $\Omega\equiv \sqrt{F^2+(\omega_0-\omega)^2}$.
(Recall that an arbitrary small amount of eccentricity $F-G$ produces an avoided level crossing
and hence a smooth maximum of the quasienergy). After some manipulations the condition  (\ref{RES7}) can be
transformed into
\begin{eqnarray}\label{RES8a}
  \frac{\omega_{res}^{(2)}}{\omega_0}&=&\frac{1}{3}\left(-1+\sqrt{3\left( \frac{F}{\omega_0}\right)^2+4}\right)\\
  \label{RES8b}
  &=&\frac{1}{3}+\sum_{n=1}^\infty \frac{(-1)^{n+1} 3^{n-1} (2 n-3)\text{!!}}{ 2^{3 n-1}\,n!}\left( \frac{F}{\omega_0}\right)^{2n}\\
  \label{RES8c}
  &=& \frac{1}{3}+\frac{1}{4} \left(\frac{F}{\omega _0}\right)^2-\frac{3}{64} \left(\frac{F}{\omega_0}\right)^4
  +\frac{9}{512} \left(\frac{F}{\omega _0}\right)^6+O\left(\frac{F}{\omega _0}\right)^8
   \;.
\end{eqnarray}
It can be easily checked that the coefficients of the power series (\ref{RES8c}) coincide with the anti-diagonal sums,
i.~e., $\tilde{\Omega}^{(0)}_2=\frac{1}{3}$, $\tilde{\Omega}^{(2)}_2=\frac{1}{4}$, $\tilde{\Omega}^{(2)}_4=-\frac{3}{64}$, and $\tilde{\Omega}^{(2)}_6=\frac{9}{512}$.

Finally, we consider the $3^{rd}$ resonance described by
\begin{equation}\label{RES9}
 \Omega^{(3)}=\left(
\begin{array}{cccccc}
 \frac{1}{5} & 0 & \frac{5}{96} & 0 & -\frac{2125}{221184} & \ldots \\
 0 & \frac{1}{48} & 0 & -\frac{125}{55296} & 0 & \ldots  \\
 \frac{5}{96} & 0 & -\frac{205}{36864} & 0 & * &\ldots  \\
 0 & -\frac{125}{55296} & 0 & * & 0 &\ldots  \\
 -\frac{2125}{221184} & 0 & * & 0 & * &\ldots  \\
\vdots& \vdots & \vdots & \vdots & \vdots & \vdots \\
\end{array}
\right)
\;,
\end{equation}
the anti-diagonal sums of which are obtained via
\begin{eqnarray}\label{RES10a}
  \frac{\omega_{res}^{(3)}}{\omega_0}&=&\frac{1}{15}\left(-1+\sqrt{15\left( \frac{F}{\omega_0}\right)^2+16}\right)\\
  \label{RES10b}
  &=&\frac{1}{5}+\sum_{n=1}^\infty \frac{(-15)^{n-1}  (-3+2 n)\text{!!}}{2^{4 n-1}\,n!}\left( \frac{F}{\omega_0}\right)^{2n}\\
  \label{RES10c}
  &=&\frac{1}{5}+\frac{1}{8} \left(\frac{F}{\omega _0}\right){}^2
-\frac{15}{512} \left(\frac{F}{\omega_0}\right)^4+O\left(\frac{F}{\omega _0}\right)^6
   \;.
\end{eqnarray}

The first non-trivial anti-diagonal of $\Omega^{(n)}$ can be given in closed form.
According to the recurrence relation given in \cite{S18} we conjecture that
\begin{equation}\label{RES9a}
  \Omega^{(n)}_{2,0}=  \Omega^{(n)}_{0,2}=\frac{2 n-1}{16 (n-1) n}
  \;,
\end{equation}
for $n>1$.
Employing the circular limit
\begin{equation}\label{RES10}
  \frac{\omega_{res}^{(n)}}{\omega_0}=\frac{\sqrt{(2 n-3) (2 n-1) \left(\frac{F^2}{\omega _0^2}+1\right)+1}-1}{4 (n-2) n+3}
  =\frac{1}{2n-1}+\frac{1}{4(n-1)}\left(\frac{F}{\omega_0}\right)^2+O\left(\frac{F}{\omega_0}\right)^4
\end{equation}
we obtain
\begin{equation}\label{RES11}
  \Omega^{(n)}_{1,1}=  \frac{1}{8 (n-1) n}
  \;,
\end{equation}
for $n>1$. These results can be checked for $n=2,3$ by inspection of (\ref{RES6}) and (\ref{RES9}).

%%%%%%%%%%%%%%%%%%%%%%%%%%%%%%%%%%%%%%%%%%%%%%%%%%%%%%%%%%%%%%%%%%%%%%%%%%%%%%%%%%%%%%%%%%%%%%%%%%%%%%%%%%%%%%%%%%%%%%%%%%%%%%
\section{Special limit cases}\label{sec:L}
%%%%%%%%%%%%%%%%%%%%%%%%%%%%%%%%%%%%%%%%%%%%%%%%%%%%%%%%%%%%%%%%%%%%%%%%%%%%%%%%%%%%%%%%%%%%%%%%%%%%%%%%%%%%%%%%%%%%%%%%%%%%%%%%%%%%%%

%%%%%%%%%%%%%%%%%%%%%%%%%%%%%%%%%%%%%%%%%%%%%%%%%%%%%%%%%%%%%%%%%%%%%%%%%%%%%%%%%%%%%%%%%%%%%%%%%%%%%%%%%%%%%%%%%%%%%%%%%%%%%%
\subsection{Limit case $\omega\rightarrow 0$}\label{sec:LOM}
%%%%%%%%%%%%%%%%%%%%%%%%%%%%%%%%%%%%%%%%%%%%%%%%%%%%%%%%%%%%%%%%%%%%%%%%%%%%%%%%%%%%%%%%%%%%%%%%%%%%%%%%%%%%%%%%%%%%%%%%%%%%%%%%%%%%%%

\begin{figure}[h]
  \centering
  \includegraphics[width=0.75\linewidth]{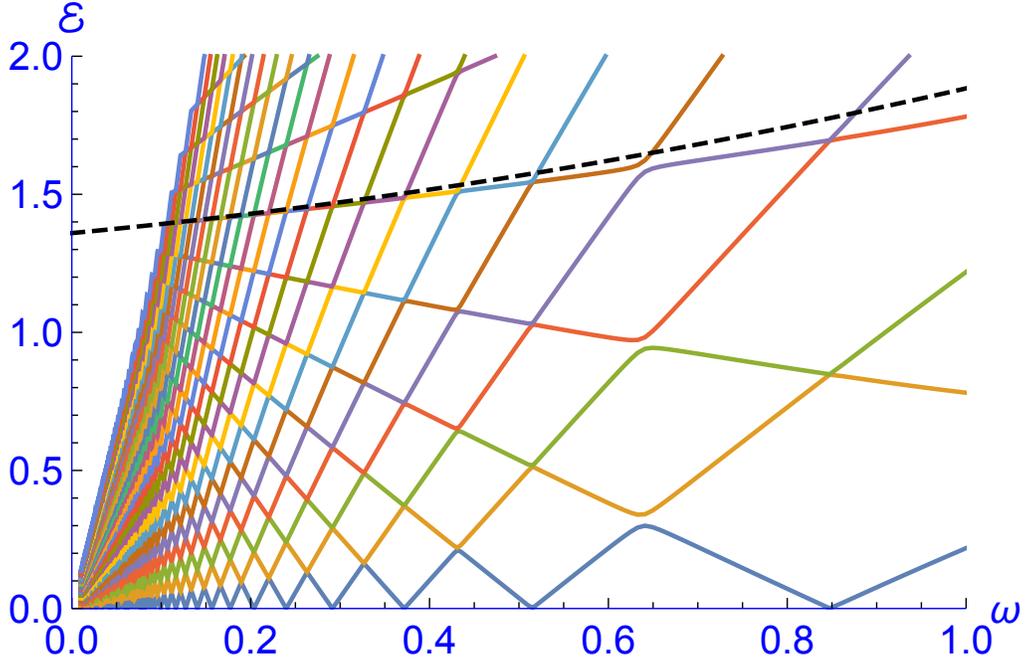}
  \caption[T1]
  {Various branches of the quasienergy ${\mathcal E}(\omega_0,F,G,\omega)$ as a function of $\omega$ for fixed values of $\omega_0=1$, $F=3$ and $G=2$.
  The different branches are generated by adding integer multiples of $\omega$ to $\pm {\mathcal E}$ and can be distinguished by their color.
  The black dashed curve represents the adiabatic approximation ${\mathcal E}_0+\omega\,{\mathcal E}_1+\omega^2\,{\mathcal E}_2$
  according to (\ref{LOM2d}), (\ref{LOM5}) and (\ref{LOM8}).
   }
  \label{FIGSEPS2}
\end{figure}

This limit case (``adiabatic limit") has been already treated in \cite{S18} in sufficient generality, such that we only need to recall the
essential issues. We adopt a series representation
\begin{equation}\label{LOM1}
 {\mathbf S}(\omega t)=\sum_{n=0}^{\infty}\omega^n\,{\mathbf S}^{(n)}(\omega t)
\end{equation}
of the periodic solution of (\ref{D1}) and obtain a recursive system of inhomogeneous linear differential equations for the ${\mathbf S}_n$.
The starting point is
\begin{equation}\label{LOM2}
  {\mathbf S}^{(0)}(\omega t)=\frac{{\mathbf h}(t)}{\left\|{\mathbf h}(t)\right\|}=
  \frac{1}
  {\sqrt{F^2 \sin^2(\omega t )+G^2 \cos ^2(\omega t )+\omega _0^2}}
  \left(\begin{array}{c}
            \omega _0 \\
          G\,\cos\omega t \\
          F\,\sin\omega t
        \end{array}
  \right)
   \;,
\end{equation}
that is, the spin vector follows the direction of the slowly varying magnetic field.
The corresponding zeroth term of the series for the quasienergy
\begin{equation}\label{LOM2a}
 {\mathcal E}= \overline{\frac{1}{2}\left({\mathbf h}_1+\frac{{\mathbf h}_2{\mathbf S}_2+{\mathbf h}_3{\mathbf S}_3}{1+{\mathbf S}_1} \right)}
=\sum_{n=0}^{\infty} {\mathcal E}_n\,\omega^n
\end{equation}
can be obtained as
\begin{eqnarray}
\label{LOM2b}
 {\mathcal E}_0 &=& \overline{\frac{1}{2}\left({\mathbf h}_1+\frac{{\mathbf h}_2{\mathbf S}^{(0)}_2+{\mathbf h}_3{\mathbf S}^{(0)}_3}{1+{\mathbf S}^{(0)}_1} \right)}
 =\overline{\frac{1}{2}\sqrt{{\mathbf h }\cdot{\mathbf h}}}
 \\
 \label{LOM2c}
   &=& \overline{\frac{1}{2} \sqrt{F^2 \sin ^2(\omega t )+G^2 \cos ^2(\omega t )+\omega _0^2}} \\
  \label{LOM2d}
   &=& \frac{\sqrt{G^2+\omega _0^2}}{\pi } E\left(\frac{G^2-F^2}{G^2+\omega _0^2}\right)
   \;,
\end{eqnarray}
where $E(\ldots)$ denoted the complete elliptic integral of the $2^{nd }$ kind.
Note that in the adiabatic limit the quasienergy $ {\mathcal E}_0$ can be completely reduced to its dynamical part ${\mathcal E}_d$,
since the geometrical part ${\mathcal E}_g$ is proportional to $\omega$ and only contributes to the next term $ {\mathcal E}_1$.
For $G=0$ the formula for ${\mathcal E}_0$ agrees with Eq.~(253) in \cite{S18}. In the circular case ($G=F$) the series expansion
\begin{equation}\label{LOM2e}
  {\mathcal E}=\frac{\omega+\Omega}{2}=\frac{1}{2} \sqrt{F^2+\omega _0^2}+\left(\frac{1}{2}-\frac{\omega _0}{2
   \sqrt{F^2+\omega _0^2}}\right) \omega +\frac{F^2 \omega ^2}{4 \left(F^2+\omega
   _0^2\right){}^{3/2}}+O\left(\omega ^3\right)
\end{equation}
yields the zeroth order contribution
$\lim_{\omega\rightarrow 0}{\mathcal E}=\lim_{\omega\rightarrow 0}\frac{\omega+\Omega}{2}=\frac{1}{2}\sqrt{F^2+\omega_0^2}$  that also follows
from (\ref{LOM2d}) and $E(0)=\frac{\pi}{2}$.\\

The next term ${\mathbf S}^{(1)}$ of the series (\ref{LOM1}) is obtained as the solution of
\begin{equation}\label{LOM3}
\frac{d}{dt} {\mathbf S}^{(0)} = {\mathbf h}\times {\mathbf S}^{(1)}
\;,
\end{equation}
such that ${\mathbf S}^{(0)}\cdot{\mathbf S}^{(1)}=0$ in order to guarantee normalization in linear $\omega$-order.
The result is
\begin{equation}\label{LOM4}
\omega\,{\mathbf S}^{(1)}(t)=\left( \frac{d}{dt}{\mathbf S}^{(0)}(t)\right)\times\frac{{\mathbf h}(t)}{\left\|{\mathbf h}(t)\right\|^2}=
\frac{2 \sqrt{2}\,\omega }{\left(\left(G^2-F^2\right) \cos (2 \omega t )+F^2+G^2+2 \omega
   _0^2\right)^{3/2}}
   \left(
\begin{array}{c}
 -F G \\
 F \, \omega _0\,\cos  \omega t \\
 G\, \omega _0\, \sin  \omega t \\
\end{array}
\right)
\;.
\end{equation}
It leads to a linear contribution to the quasienergy of
\begin{equation}\label{LOM5}
 \omega\,{\mathcal E}_1=\omega\left(
 \frac{1}{2}-\frac{F \omega _0 \Pi \left(1-\frac{F^2}{G^2}|\frac{G^2-F^2}{G^2+\omega_0^2}\right)}{\pi  G \sqrt{G^2+\omega _0^2}}\right)
   \;,
\end{equation}
where $\Pi(\ldots|\ldots)$ denotes the complete elliptic integral of the $3^{rd}$ kind. According to (\ref{LOM4}),
${\mathbf S}_1\cdot {\mathbf h}=0$ and hence the dynamical part ${\mathcal E}_{1d}$ of ${\mathcal E}_1$ vanishes.
${\mathcal E}_1$ consists only of the geometrical part that can be identified with the Berry phase
\cite{B84}, \cite{AA87}, \cite{MB14} divided by the period
since in the adiabatic limit the solid angles swept by the elliptically polarized magnetic field and by the spin vector are identical.
Consequently, ${\mathcal E}_1$ vanishes in the limit of linear polarization. For $G=F$ the limit of (\ref{LOM5}) and
the linear term in the series expansion (\ref{LOM2e}) agree since $\Pi(0,0)=\frac{\pi}{2}$.

According to \cite{S18}, the next, quadratic term of (\ref{LOM1}) is given by
\begin{eqnarray}\label{LOM6a}
{\mathbf S}^{(2)}&=&\left(\frac{1}{\omega}\frac{d}{dt}{\mathbf S}^{(1)}\right)\times\frac{\mathbf h}{\|{\mathbf h}\|^2}
-\frac{\mathbf h}{2\|{\mathbf h}\|^2}{\mathbf S}^{(1)}\cdot{\mathbf S}^{(1)}\\
\label{LOM6b}
&=&\frac{1}{\left(\left(G^2-F^2\right) \cos (2 t \omega )+F^2+G^2+2 \omega_0^2\right)^{7/2}}
\left(
\begin{array}{c}
 {\mathbf S}^{(2)}_{1,0}+{\mathbf S}^{(2)}_{1,2}\cos(2\omega t)+{\mathbf S}^{(2)}_{1,4}\cos(4\omega t) \\
  {\mathbf S}^{(2)}_{2,1}\cos(\omega t)+{\mathbf S}^{(2)}_{2,3}\cos(3\omega t) \\
  {\mathbf S}^{(2)}_{3,1}\sin(\omega t)+{\mathbf S}^{(2)}_{3,3}\sin(3\omega t)
\end{array}
\right)
\;,
\end{eqnarray}
where
\begin{eqnarray}
\label{LOM7a}
 {\mathbf S}^{(2)}_{1,0} &=& -6 \sqrt{2} \omega _0 \left(F^4+\omega _0^2 \left(F^2+G^2\right)+G^4\right),\\
 \label{LOM7b}
  {\mathbf S}^{(2)}_{1,2} &=& 2 \sqrt{2} \omega _0 \left(F^2-G^2\right) \left(2 F^2+2 G^2+\omega _0^2\right), \\
 \label{LOM7c}
  {\mathbf S}^{(2)}_{1,4} &=& 2 \sqrt{2} \omega _0 \left(F^2-G^2\right)^2,  \\
 \label{LOM7d}
  {\mathbf S}^{(2)}_{2,1} &=&  \sqrt{2} G \left(-6 F^4+F^2 \left(2 G^2-7 \omega _0^2\right)+11 G^2 \omega _0^2+8 \omega_0^4\right),\\
 \label{LOM7e}
  {\mathbf S}^{(2)}_{2,3} &=& 3 \sqrt{2} G (F-G) (F+G) \left(2 F^2+\omega _0^2\right),\\
 \label{LOM7f}
  {\mathbf S}^{(2)}_{3,1} &=& \sqrt{2} F \left(F^2 \left(2 G^2+11 \omega _0^2\right)-6 G^4-7 G^2 \omega _0^2+8 \omega_0^4\right),\\
  \label{LOM7g}
  {\mathbf S}^{(2)}_{3,3} &=&3 \sqrt{2} F \left(F^2-G^2\right) \left(2 G^2+\omega _0^2\right)
  \;.
\end{eqnarray}

The corresponding quadratic correction to the quasienergy is too complicated to be calculated here.
We confine ourselves to determine ${\mathcal E}_2$ for a special set of physical parameters, namely
$F=3$, $G=2$, and $\omega_0=1$. The result is
\begin{equation}\label{LOM8}
 {\mathcal E}_2=57-\frac{16147}{200 \sqrt{2}}+
 \frac{1}{\pi}\left(
 \frac{4777\, \Gamma \left(\frac{1}{4}\right)^2+21036\, \Gamma
   \left(\frac{3}{4}\right)^2}{240 \sqrt{10 \pi }}-\frac{171\, \Pi
   \left(\left.-\frac{5}{4}\right|-1\right)}{\sqrt{5}}\right)
   \approx 0.217319\ldots
   \;.
\end{equation}

The corresponding adiabatic approximation of the quasienery has been shown in Figure \ref{FIGSEPS2}
together with the various branches of the form $n\,\omega\pm {\mathcal E}$. It turns out that the
adiabatic approximation is a kind of envelope of a certain family of branches that interpolates between the
numerous avoided level crossings of this family.
This finding is insofar plausible, since by definition the adiabatic limit of quasi-energy is an
analytical function of $\omega$, while the different branches $n\,\omega\pm {\mathcal E}$ for
$\omega\rightarrow 0$ get stronger and stronger kinks.

%%%%%%%%%%%%%%%%%%%%%%%%%%%%%%%%%%%%%%%%%%%%%%%%%%%%%%%%%%%%%%%%%%%%%%%%%%%%%%%%%%%%%%%%%%%%%%%%%%%%%%%%%%%%%%%%%%%%%%%%%%%%%%%
\subsection{Limit case $F, G\rightarrow 0$}\label{sec:FG}
%%%%%%%%%%%%%%%%%%%%%%%%%%%%%%%%%%%%%%%%%%%%%%%%%%%%%%%%%%%%%%%%%%%%%%%%%%%%%%%%%%%%%%%%%%%%%%%%%%%%%%%%%%%%%%%%%%%%%%%%%%%%%%

For sake of comparison with the analogous results in \cite{S18} we rewrite the equation of motion (\ref{D1})
in the form
\begin{eqnarray}
\label{FG1a}
  \frac{d X}{dt} &=&\lambda\,G\,\cos(\omega t)\,Z-\lambda\,F\,\sin(\omega t)\,Y\;, \\
\label{FG1b}
  \frac{d Y}{dt} &=&\lambda\,F\,\sin(\omega t)\,X-\omega_0\,Z\;, \\
\label{FG1c}
  \frac{d Z}{dt} &=&\omega_0\,Y-\lambda\,G\,\cos(\omega t)\,X
  \;,
\end{eqnarray}
where $\lambda$ is a formal expansion parameter that is ultimately set to $\lambda=1$.

In the case $\lambda=0$ there are only two normalized solutions of the classical Rabi problem
that are  $T$-periodic for all $T>0$,
namely ${\mathbf X}(t)=\pm (1,0,0)^\top$. Hence for infinitesimal $\lambda$ we expect that we still have
$X(t)=\pm 1+O(\lambda^2)$ but $(Y(t),Z(t))$ will describe an infinitesimal ellipse, i.~e.~,
$Y(t)= A \cos\omega t+O(\lambda^3)$ and $Z(t)=B \sin\omega t+O(\lambda^3)$, such that $A$ and $B$ depend linearly
on $\lambda F$ and $\lambda G$.
These considerations and numerical investigations suggest the following Fourier-Taylor (FT) series ansatz,
not yet normalized,
\begin{eqnarray}
\nonumber
  X(t)&=&\sum_{n=0}^{\infty}\lambda^{2n}\sum_{m=0}^{n}R_{n,m}(F,G,\omega,\omega_0)\cos\,2m\omega t\;, \\
  &&\label{FL2a}\\
 \nonumber
  Y(t) &=& \sum_{n=0}^{\infty}\lambda^{2n+1}\sum_{m=0}^{n}S_{n,m}(F,G,\omega,\omega_0)\cos\,(2m+1)\omega t\;, \\
   \label{FL2b}
  &&  \\
  \label{FL2c}
  Z(t)&=& \sum_{n=0}^{\infty}\lambda^{2n+1}\sum_{m=0}^{n}T_{n,m}(F,G,\omega,\omega_0)\sin\,(2m+1)\omega t\;.
\end{eqnarray}

Inserting these series into the differential equations (\ref{FG1a}) -- (\ref{FG1c}) and collecting powers of $\lambda$ yields recurrence relations
for the functions $R_{n,m},\,S_{n,m}$ and $T_{n,m}$.
As  initial conditions we use the following choices that result from the above considerations and
the lowest orders $\lambda^0$ and $\lambda^1$ of the differential equations (\ref{FG1a}) -- (\ref{FG1c}):
\begin{eqnarray}
\label{FL3a}
  R_{0,0}(F,G,\omega,\omega_0) &=& 1\;,\\
    \label{FL3b}
  R_{n,0}(F,G,\omega,\omega_0) &=& 0\quad \mbox{for } n=1,2,\ldots,\\
  \label{FL3c}
  S_{0,0}(F,G,\omega,\omega_0)&=& -\frac{F \omega +G \omega _0}{\left(\omega -\omega _0\right) \left(\omega +\omega _0\right)}\;,\\
   \label{FL3d}
  T_{0,0}(F,G,\omega,\omega_0)&=& -\frac{F \omega _0+G \omega }{\left(\omega -\omega _0\right) \left(\omega +\omega _0\right)}\;.
\end{eqnarray}

For $n>0$ the FT coefficients $R_{n,m},\,S_{n,m}$ and $T_{n,m}$ can be recursively determined by means of the
following relations:
\begin{eqnarray}
\nonumber
  R_{n+1,m} &=& \frac{1}{4\,m\,\omega}\left(F\,S_{n,m-1} -F\,S_{n,m}-G\,T_{n,m-1}-G\,T_{n,m}\right)\\
  \label{FL4a}
  && \mbox{ for } 1\le m\le n+1\;, \\
  \nonumber
  S_{n,m} &=& \frac{1}{2 \left(((2 m +1)\omega )^2-\omega _0^2\right)}\,
  \left(\left(-G \omega _0-F (2 m+1) \omega \right)R_{n,m} + \left( -G \omega _0+F (2   m+1) \omega \right)R_{n,m+1} \right)\\
   \label{FL4b}
   && \mbox{ for } 0\le m\le n\;,\\
   \nonumber
  T_{n,m} &=& \frac{1}{2 \left(((2 m+1) \omega )^2-\omega _0^2\right)}
  \left(
  \left(-G (1+2 m) \omega -F \omega _0\right) R_{n,m}+\left(-G (1+2 m) \omega +F \omega _0\right) R_{n,m+1}
  \right)
  \\
   && \mbox{ for } 0\le m\le n\;.
  \label{FL4c}
\end{eqnarray}
where, of course, we have to set $R_{n,n+1}=S_{n,n+1}=0$ in (\ref{FL4a}) - (\ref{FL4c}).
It follows
that $R_{n,m}(F,G,\omega,\omega_0)$, $S_{n,m}(F,G,\omega,\omega_0)$ and $T_{n,m}(F,G,\omega,\omega_0)$
are rational functions of their arguments.

%We recall that under the transformation (\ref{H2a}) -- (\ref{H2e}),  $X,Y$ and $Z$ remain invariant which entails
%$R_{nm}\mapsto \lambda^{-2n-1 }\,R_{nm}$ and $S_{nm}\mapsto \lambda^{-2n}\, S_{nm}$. Then it easily follows that both sides of the
%recurrence relations (\ref{FL4a}) and (\ref{FL4b}) transform in the same way, which can be viewed as a consistency check of the ansatz
%(\ref{FL2a}) -- (\ref{FL2c}).

We will show the first few terms of the FT series for $X(t), Y(t)$ and $Z(t)$:
\begin{eqnarray}
\nonumber
  X(t) &=&1-\frac{(F-G) (F+G)}{4 \left(\omega^2 -\omega_0^2\right)}\,\cos 2\omega t\quad + \\
  \nonumber
  &&
   -\frac{(F-G) (F+G) \left(3 F^2 \omega ^2+3 G^2 \omega ^2-4 F G \omega  \omega _0-F^2
   \omega _0^2-G^2 \omega _0^2\right)}{8 \left(\omega^2 -\omega_0^2\right)^2
   \left(9\omega^2 -\omega_0^2\right)}\,\cos 2\omega t\\
  \nonumber
  &&+ \left(
  \frac{3 (F-G)^2 (F+G)^2}{64 \left(\omega^2 -\omega_0^2\right)
   \left(9 \omega^2 -  \omega_0^2\right)}
  \right)
   \cos 4\omega t\,
    + O(\lambda^6),
    \\
    \label{FL5a}
   &&\\
   \nonumber
   Y(t)&=& \left(
   -\frac{F \omega +G \omega _0}{\left(\omega^2 -\omega_0^2\right)}
   \right)\,\cos \omega t\\
   \label{FL5b}
   &&+
   \left(
   -\frac{(F-G) (F+G) \left(F \omega -G \omega _0\right)}{8 \left(\omega^2 -\omega_0^2\right){}^2}
   \right)\,\cos 3 \omega t
    + O(\lambda^5)\;,\\
      \nonumber
   Z(t)&=& \left(
  -\frac{G \omega +F \omega _0}{\left(\omega^2 -\omega_0^2\right) }
   \right)\,\sin \omega t\\
   \label{FL5c}
   &&+
   \left(
  -\frac{(F-G) (F+G) \left(-G \omega +F \omega _0\right)}{8 \left(\omega^2 -\omega_0^2\right)^2 }
   \right)\,\sin 3 \omega t
    + O(\lambda^5)
    \;,
    \end{eqnarray}
where $\lambda$ stands for any linear combination of $F$ and $G$.
We note that the coefficients
contain denominators of the form $\omega^2-\omega_0^2$ and $9\omega^2-\omega_0^2$ due to the denominator
$(2m+1)^2\omega^2-\omega_0^2$ in the recursion relations (\ref{FL4b}) and (\ref{FL4c}).
Hence the FT series breaks down at the
resonance frequencies $\omega_{res}^{(m)}=\frac{\omega_0}{2m-1}$.
This is the more plausible since
according to the above ansatz $z_0=1$ which is not compatible with the resonance condition $z_0=0$ mentioned above.

Using the FT series solution (\ref{FL2a}) -- (\ref{FL2c}) it is a straightforward task to calculate the quasienergy
${\mathcal E}=a_0$ as the time-independent part of the FT series of
\begin{equation}\label{FL6}
\frac{1}{2}\left(\omega_0+\frac{G \cos(\omega t)Y(t)+F \sin(\omega t)Z(t)}{R+Z(t)}\right)
=a_0+\sum_{\stackrel{n\in{\mathbbm Z}}{n\neq 0}}a_n\,e^{{\sf i}\,n\,\omega\, t}\;,
\end{equation}
according to (\ref{D8}).
The first few terms of the result are given by
\begin{eqnarray}
\nonumber
{\mathcal E} &=& \frac{\omega_0}{2}
-\frac{2 F G \omega +F^2 \omega _0+G^2 \omega _0}{8 \left(\omega ^2-\omega_0^2\right)} \\
  \nonumber
   &&+\frac{4 F G \left(F^2+G^2\right) \omega ^3+\left(F^4+22 F^2 G^2+G^4\right) \omega ^2
   \omega _0+12 F G \left(F^2+G^2\right) \omega  \omega _0^2+\left(3 F^4+2 F^2 G^2+3
   G^4\right) \omega _0^3}{128 \left(\omega ^2-\omega _0^2\right){}^3}
   + O(\lambda^6)\;.\\
    \label{FL7}
   &&
\end{eqnarray}
This is in agreement with the result for linear polarization, see \cite{S18}, eq.~(198), if we set $G=0$.\\

It will be instructive to check the first two terms of (\ref{FL7}) by using the decomposition of the quasienergy into
a dynamical and a geometrical part. In lowest order in $\lambda$ the classical RPE solution is a motion on
an ellipse with semi axes
\begin{equation}\label{FL8}
a=\frac{F \omega +G \omega _0}{\left|\omega^2 -\omega_0^2\right|}\;,
\quad\mbox{and }
b=\frac{G \omega +F \omega _0}{\left|\omega^2 -\omega_0^2\right|}
\;.
\end{equation}
Hence the geometrical part of the quasienergy reads
\begin{equation}\label{FL9}
 {\mathcal E}_g=\frac{\omega}{4\pi}\,\pi\,a\,b+O\left(\lambda^4\right)
 = \frac{\omega  \left(G \omega +F \omega _0\right) \left(F \omega +G \omega _0\right)}
 {4\left(\omega ^2-\omega _0^2\right)^2}+O(\lambda^4)\;.
\end{equation}
The dynamical part is obtained as
\begin{equation}\label{FL10}
 {\mathcal E}_d=\overline{\frac{\omega_0\,X+ G \,\cos(\omega t)\,Y+F \,\sin(\omega t)\,Z}{2 R}}
   =\frac{\omega_0}{2}
   +\frac{-4 F G \omega ^3-3 \left(F^2+G^2\right) \omega^2 \omega_0+\left(F^2+G^2\right)
   \omega_0^3}{8 \left(\omega^2-\omega _0^2\right)^2}
   +O\left(\lambda^4\right).
\end{equation}
The sum of both parts together correctly yields
\begin{equation}\label{FL11}
{\mathcal E}={\mathcal E}_d + {\mathcal E}_g= \frac{\omega_0}{2}-\frac{2 F G \omega
+\left(F^2+G^2\right) \omega _0}{8 \left(\omega^2-\omega_0^2\right)}
+O\left(\lambda^4\right)\;.
\end{equation}
Moreover, the slope relation (\ref{QIIIslope}) is satisfied in the considered order,
\begin{equation}\label{FL12}
\frac{\partial {\mathcal E}}{\partial \omega} =
\frac{\left(G \omega +F \omega _0\right) \left(F \omega +G \omega _0\right)}{4\left(\omega ^2-\omega _0^2\right){}^2}+O(\lambda^4)
=\frac{ {\mathcal E}_g}{\omega}
\;,
\end{equation}
in accordance with \cite{S18}, eq.~(202).

However, as mentioned above, the FT series for the quasienergy has poles at the values $\omega=\omega_{res}^{(m)} = \frac{1}{2m-1},\;m=1,2,\ldots$
and hence the present FT series ansatz is not suited to investigate the Bloch-Siegert shift for small $\lambda$.
We have thus chosen another approach in Section \ref{sec:RES}.

%%%%%%%%%%%%%%%%%%%%%%%%%%%%%%%%%%%%%%%%%%%%%%%%%%%%%%%%%%%%%%%%%%%%%%%%%%%%%%%%%%%%%%%%%%%%%%%%%%%%%%%%%%%%%%%%%%%%%%%%%%%%%%%
\subsection{Limit case $\omega_0\rightarrow 0$}\label{sec:LOMO}
%%%%%%%%%%%%%%%%%%%%%%%%%%%%%%%%%%%%%%%%%%%%%%%%%%%%%%%%%%%%%%%%%%%%%%%%%%%%%%%%%%%%%%%%%%%%%%%%%%%%%%%%%%%%%%%%%%%%%%%%%%%%%%

It is well-known, see, e.~g., \cite{S18} or \cite{SSH20a},
that for $\omega_0=0$ and linear polarization ($F=0$) the equation of motion (\ref{FG1a}) - (\ref{FG1a})
has the exact solution
\begin{eqnarray}\label{LOMO1a}
 X(t)&=&\cos \left(\frac{G}{\omega}\, \sin \,\omega  t\right)
 =J_0\left(\frac{G}{\omega}\right)+2 \sum_{m=1}^{\infty}J_{2m}\left(\frac{G}{\omega}\right)\,\cos\,2m\omega t
 \;,\\
 \label{LOMO1b}
 Y(t)&=&0\;,\\
 \label{LOMO1c}
 Z(t)&=&- \sin \left(\frac{G}{\omega}\, \sin \,\omega  t\right)
 = -2\sum_{n=0}^{\infty}J_{2m+1}\left(\frac{G}{\omega}\right)\,\sin\,(2m+1)\omega t
 \;,
\end{eqnarray}
where the $J_k(\ldots)$ denote the Bessel functions of the first kind and the series representation results
from the Jacobi-Anger expansion. Upon inserting the Taylor series of  $J_k(x)$, that starts with the lowest power $x^k$,
into (\ref{LOMO1a}) and (\ref{LOMO1c}) we would obtain the Fourier-Taylor (FT) series of $X(t)$ and $Z(t)$.
On the other hand, we have considered an FT series of $X(t), Y(t)$ and $Z(t)$ in section \ref{sec:FG} that can be specialized
to $\omega_0=F=0$. (We will indicate the specialization to  $\omega_0=F=0$ by using the notation ${\sf X}(t), {\sf Y}(t)$ and ${\sf Z}(t)$.)
The only difference is normalization: The solution (\ref{LOMO1a}) - (\ref{LOMO1c}) is already normalized
and satisfies $\overline{{\sf X}(t)}=J_0\left(\frac{G}{\omega}\right)$,
whereas the ansatz (\ref{FL2a}) - (\ref{FL2c}) assumes $\overline{{\sf X}(t)}=1$. It follows that
the FT series  (\ref{FL2a}) - (\ref{FL2c}), specialized to $\omega_0=F=0$ is identical with the FT series
obtained by (\ref{LOMO1a}) - (\ref{LOMO1c}) upon division by $J_0\left(\frac{G}{\omega}\right)$.
We have checked this for a couple of examples. Especially, it follows that
\begin{equation}\label{LOMO2}
  {\sf R}_{n,m}(g)=\left[\frac{2 J_{2 m}(g)}{J_0(g)} \right]_{2n}\,g^{2n}
  \;,
\end{equation}
where $g\equiv \frac{G}{\omega}$ and $[f(x)]_n$ denotes the coefficient $a_n$ of the Taylor series $f(x)=\sum_n a_n\,x^n$.

Unfortunately, it does not seem possible to generalize the above $\omega_0=0$ solution obtained for the linear polarization case to the elliptical case.
However, its FT series is already known: We have only to specialize (\ref{FL2a}) - (\ref{FL2c}) to the case $\omega_0=0$.
But unlike in the case of $\omega_0=F=0$, the summation over $n$ involved in this FT series cannot be performed to obtain a result in closed form.

%%%%%%%%%%%%%%%%%%%%%%%%%%%%%%%%%%%%%%%%%%%%%%%%%%%%%%%%%%%%%%%%%%%%%%%%%%%%%%%%%%%%%%%%%%%%%%%%%%%%%%%%%%%%%%%%%%%%%%%%%%%%%%%
\subsubsection{Limit case $\omega_0=0$ and $F\rightarrow 0$}\label{sec:LOMOF}
%%%%%%%%%%%%%%%%%%%%%%%%%%%%%%%%%%%%%%%%%%%%%%%%%%%%%%%%%%%%%%%%%%%%%%%%%%%%%%%%%%%%%%%%%%%%%%%%%%%%%%%%%%%%%%%%%%%%%%%%%%%%%%

\begin{figure}[h]
  \centering
  \includegraphics[width=0.75\linewidth]{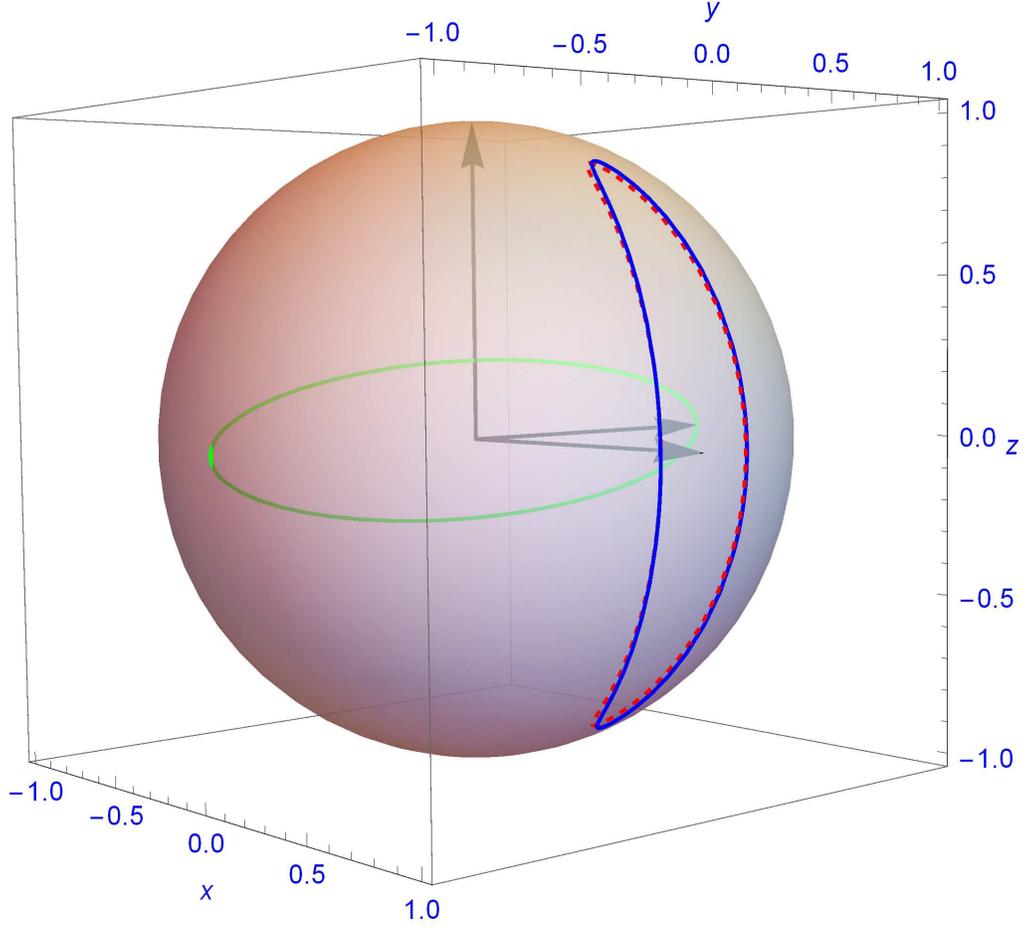}
  \caption[T1]
  {The periodic solution of the equation of motion (\ref{FG1a}) - (\ref{FG1a}) with the values of the parameters $F=1/4,\; G=\omega=1,\;  \omega_0=0$
  according to numerical integration (blue curve) and analytical approximation (\ref{LOMO1a}), (\ref{LOMO5}) and (\ref{LOMO1c}) (dashed red curve).
  The green curve represents the ellipse in the $y-z$-plane swept by the magnetic field vector.
   }
  \label{FIGXYZ}
\end{figure}

We can only get a result for ``almost linear" polarization, i.~e., in the lowest linear order of $F$. To achieve this result we
first note that for $\omega_0=0$ the functions $X(t)$ and $Z(t)$ will be even functions of $F$ and $Y(t)$ will be an odd one.
This is compatible with the above-mentioned fact that ${\sf Y}(t)$ vanishes for $F=0$ and can be shown by induction over $n$ using
the recurrence relations  (\ref{FL4a}) -  (\ref{FL4c}). It follows that the linear part $Y_1(t)$ of $Y(t)=Y_1(t)\, F + Y_3(t)\,F^3+\ldots$
can be obtained by applying the recurrence relation  $(\ref{FL4b})$ that reduces to
\begin{eqnarray}\label{LOMO3a}
 S_{n,m}&=&\frac{F}{2 ((2 m+1) \omega )}\left( R_{n,m+1}-R_{n,m}\right)\\
 \label{LOMO3b}
 &\stackrel{(\ref{LOMO2})}{=}& \frac{F g^{2 n}}{2 ((2 m+1) \omega )}\left( \left[\frac{2 J_{2( m+1)}(g)}{J_0(g)} \right]_{2n}
 -\left[\frac{2 J_{2 m}(g)}{J_0(g)} \right]_{2n}\right)
 \;.
\end{eqnarray}
Now we can perform the sum over $n=0,\ldots,\infty$ without any problems:
\begin{eqnarray}
\label{LOMO4a}
  \sum_{n=0}^{\infty}S_{n,m} &=&  \sum_{n=0}^{\infty}\frac{F g^{2 n}}{2 ((2 m+1) \omega )}\left( \left[\frac{2 J_{2 (m+1)}(g)}{J_0(g)} \right]_{2n}
 -\left[\frac{2 J_{2 m}(g)}{J_0(g)} \right]_{2n}\right) \\
   &=& \frac{F (J_{2 (m+1)}(g)-J_{2 m}(g))}{(1+2 m) \omega  J_0(g)}
   \;,
\end{eqnarray}
which, finally, yields
\begin{equation}\label{LOMO5}
Y(t)=Y_1(t)F+O(F^3)= \frac{F}{\omega}\,\sum_{m=0}^{\infty}\frac{J_{2 (m+1)}(g)-J_{2 m}(g)}{(1+2 m)}\,\cos (2m+1)\omega t +O(F^3)
\;,
\end{equation}
where we have multiplied the result by $ J_0(g)$ in order to obtain a normalized solution.
The analytical approximation given by  (\ref{LOMO1a}), (\ref{LOMO5}) and (\ref{LOMO1c}) is surprisingly of good quality even for
relative large values of, say,  $\frac{F}{G}\sim 1/4$, see Figure \ref{FIGXYZ}.

%%%%%%%%%%%%%%%%%%%%%%%%%%%%%%%%%%%%%%%%%%%%%%%%%%%%%%%%%%%%%%%%%%%%%%%%%%%%%%%%%%%%%%%%%%%%%%%%%%%%%%%%%%%%%%%%%%%%%%%%%%%%%%%
\subsubsection{Limit case $\omega_0=0$ and $F\rightarrow G$}\label{sec:LOMOC}
%%%%%%%%%%%%%%%%%%%%%%%%%%%%%%%%%%%%%%%%%%%%%%%%%%%%%%%%%%%%%%%%%%%%%%%%%%%%%%%%%%%%%%%%%%%%%%%%%%%%%%%%%%%%%%%%%%%%%%%%%%%%%%

\begin{figure}[h]
  \centering
  \includegraphics[width=0.75\linewidth]{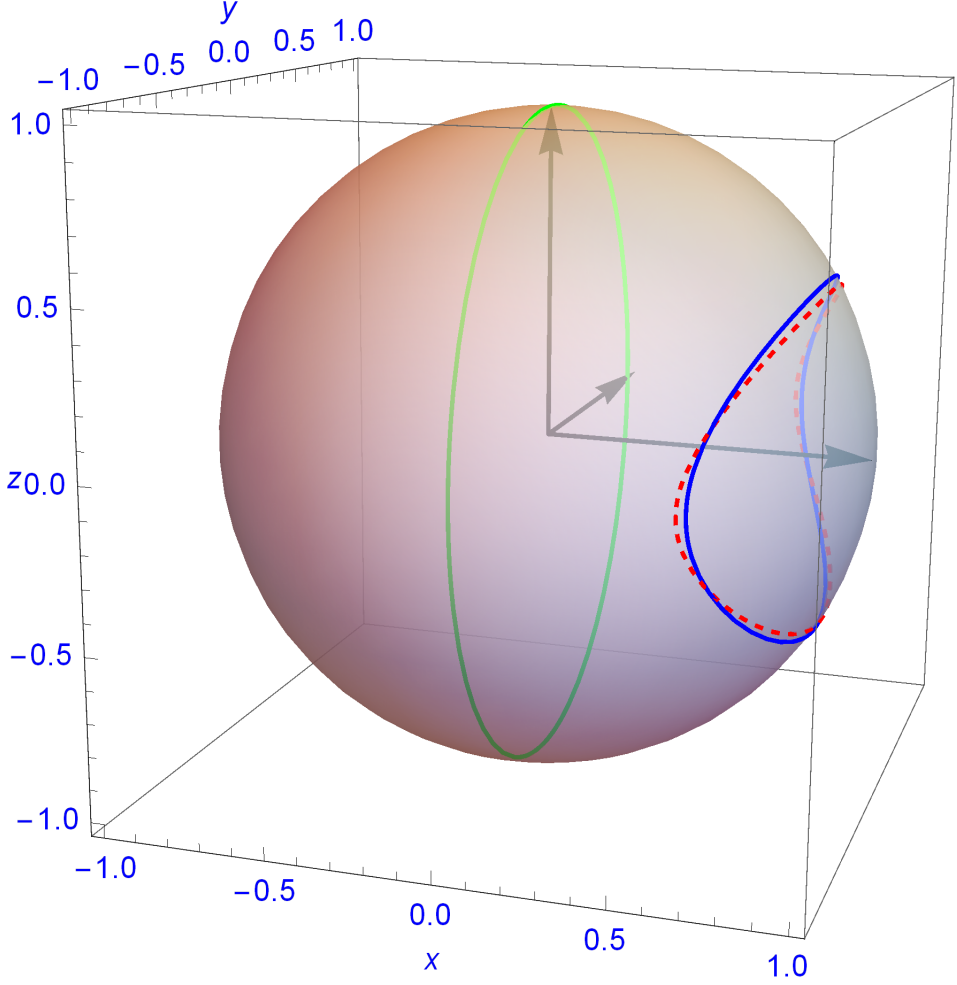}
  \caption[T1]
  {The periodic solution of the equation of motion (\ref{FG1a}) - (\ref{FG1a}) with the values of the parameters $F=1,\; G=3/4,\;\omega=1,\;  \omega_0=0$
  according to numerical integration (blue curve) and (normalized) analytical approximations (\ref{LOMOC6a}) - (\ref{LOMOC6c}) (dashed red curve).
  The green curve represents the ellipse in the $y-z$-plane swept by the magnetic field vector.
   }
  \label{FIGC1}
\end{figure}

The Rabi problem with circular polarization ($F=G$) has two simple periodic (not yet normalized) solutions of (\ref{FG1a}) - (\ref{FG1c}), namely
\begin{equation}\label{LOMOC1}
  \left( \begin{array}{c}
           X(t) \\
           Y(t) \\
           Z(t)
         \end{array}
  \right)=
  \pm
   \left( \begin{array}{c}
          \omega-\omega_0 \\
        -  F \cos\omega t \\
        - F \sin\omega t
         \end{array}
  \right)
  \;,
\end{equation}
see, e.~g., \cite{S18}, eq.~(69). Let us consider the special solution for $\omega_0=0$
\begin{equation}\label{LOMOC2}
  \left( \begin{array}{c}
           X_c(t) \\
           Y_c(t) \\
           Z_c(t)
         \end{array}
  \right)=
     \left( \begin{array}{c}
        1 \\
        - \frac{F}{\omega} \cos\omega t \\
        - \frac{F}{\omega} \sin\omega t
         \end{array}
  \right)
  \;,
\end{equation}
and look for corrections in linear order of the parameter $\delta$ describing eccentricity, namely
\begin{equation}\label{LOMOC3}
  \delta\equiv F-G
  \;.
\end{equation}
To this end we insert $\omega_0=0$ and $G=F-\delta$ into the FT series solution (\ref{FL2a}) -  (\ref{FL2c})
and expand the FT series coefficients up to terms linear in $\delta$.
The $\delta=0$ parts of the coefficients satisfy
\begin{eqnarray}
\label{LOMOC4a}
  R_{0,0} &=& 1\;, \\
  \label{LOMOC4b}
  S_{0,0} &=& -\frac{F}{\omega}\;, \\
  \label{LOMOC4c}
  T_{0,0} &=& -\frac{F}{\omega}\;,
\end{eqnarray}
in accordance with (\ref{LOMOC2}). The $\delta$-linear parts are given by
\begin{eqnarray}
\label{LOMOC5a}
  R_{n,1} &=& -\frac{3^{1-n}}{2}\frac{F^{2 n-1}}{ \omega ^{2 n}}\,\delta,\quad\mbox{for }n=1,2,\ldots\;, \\
  \label{LOMOC5b}
  S_{n,0} &=&-\frac{3^{1-n}}{4}\frac{F^{2 n}}{ \omega ^{2 n-1}}\,\delta,\quad\mbox{for }n=1,2,\ldots\;, \\
  \label{LOMOC5c}
  T_{0,0} &=&\frac{\delta}{\omega}\;, \\
   \label{LOMOC5d}
  T_{n,0} &=&\frac{3^{1-n}}{4}\frac{F^{2 n}}{ \omega ^{2 n+1}}\,\delta,\quad\mbox{for }n=1,2,\ldots\;, \\
   \label{LOMOC5e}
  S_{n,1} &=&\frac{3^{-n}}{4}\frac{F^{2 n}}{ \omega ^{2 n+1}}\,\delta,\quad\mbox{for }n=1,2,\ldots\;.
\end{eqnarray}
It is straightforward to perform the summations over $n$ and to insert the results into  (\ref{FL2a}) -  (\ref{FL2c})
thus obtaining the analytical approximations
\begin{eqnarray}
\label{LOMOC6a}
  X_a(t) &=&1+\frac{3 \delta  F }{2 \left(F^2-3 \omega ^2\right)}\cos (2 \omega t )\;, \\
  \label{LOMOC6b}
  Y_a(t) &=& \left(-\frac{F}{\omega }+\frac{3 \delta  F^2}{4 \omega  \left(F^2-3 \omega ^2\right)}\right) \cos (\omega t )-\frac{\delta  F^2 }
  {4 \omega \left(F^2-3 \omega ^2\right)}\cos (3 \omega t)\;,\\
  \label{LOMOC6c}
  Z_a(t) &=& \left(-\frac{F}{\omega }+\frac{\delta}{F} \left(\frac{1}{\omega }-\frac{3 F^2}
  {4  \omega (F^2- 3 \omega^2)}\right)\right) \sin (\omega t)-\frac{\delta  F^2 }{4 \omega  \left(F^2-3 \omega ^2\right)} \sin (3 \omega t )
   \;.
\end{eqnarray}
The quality of these approximations is surprisingly high, see Figure \ref{FIGC1}, where a  deviation between analytical
approximation and numerical integration is only visible for $\delta \sim \frac{1}{4}$.

%%%%%%%%%%%%%%%%%%%%%%%%%%%%%%%%%%%%%%%%%%%%%%%%%%%%%%%%%%%%%%%%%%%%%%%%%%%%%%%%%%%%%%%%%%%%%%%%%%%%%%%%%%%%%%%%%%%%%%%%%%%%%%
\section{Application: Work performed on a two level system}\label{sec:WO}
%%%%%%%%%%%%%%%%%%%%%%%%%%%%%%%%%%%%%%%%%%%%%%%%%%%%%%%%%%%%%%%%%%%%%%%%%%%%%%%%%%%%%%%%%%%%%%%%%%%%%%%%%%%%%%%%%%%%%%%%%%%%%%%%%%%%%%

As an application of the results obtained in the preceding sections  we consider the work performed on a two level system
by an elliptically polarized magnetic field during one period. For a related experiment see \cite{N18}.
In contrast to classical physics this work is not just a number but,
following \cite{TLH07}, has to be understood in terms of two subsequent energy measurements.
At the time $\tau=0$ the two level system is assumed to be in a mixed state according to the canonical ensemble
\begin{equation}\label{W1}
 W=\exp\left(-\beta H(0)\right)/\mbox{Tr}\left(\exp\left(-\beta H(0)\right)\right)
 \;,
 \end{equation}
 with dimensionless inverse temperature $\beta=\frac{\hbar \,\omega}{k_B\,T}$ and
 \begin{equation}\label{W1a}
  H(0)=\frac{\nu}{2}\left(
  \begin{array}{cc}
    0 & 1\\
    1&0
  \end{array}
  \right)
  \;.
 \end{equation}
Then  at the time $\tau=0$ one performs a L\"uders measurement of the instantaneous energy $H(0)$ with the two possible
outcomes $\pm\frac{\nu}{2}$.
Hence after the measurement
the system is in the pure state $P_1$ with probability $\mbox{Tr}\left( P_1 W\right)=\frac{1}{Z}e^{-\beta\nu/2}$
or in the pure state $P_2$ with probability $\mbox{Tr}\left( P_2 W\right)=\frac{1}{Z}e^{\beta\nu/2}$, where
$P_1$ and $P_2$ are the projectors onto the eigenstates of $H(0)$, i.~e.,
\begin{equation}\label{W2}
 P_1=\frac{1}{2}\left(
  \begin{array}{cc}
    1 & 1\\
    1&1
  \end{array}
  \right)
  \;,
  \quad
   P_2=\frac{1}{2}\left(
  \begin{array}{rr}
    1 & -1\\
    -1&1
  \end{array}
  \right)
  \;,
\end{equation}
and $Z=e^{-\beta\nu/2}+e^{\beta\nu/2}$.

\begin{figure}[h]
  \centering
  \includegraphics[width=1.0\linewidth]{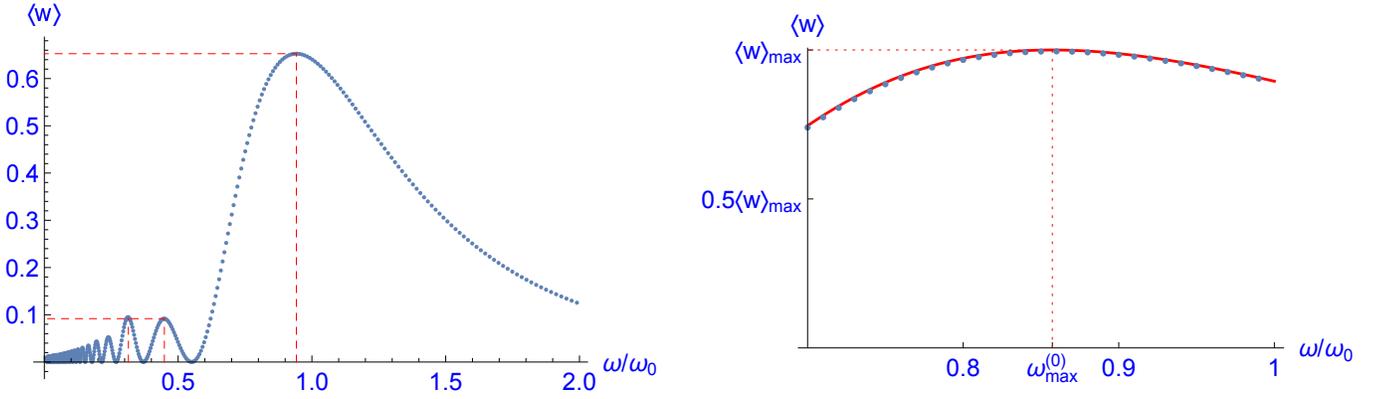}
  \caption[W1]
  {The mean value $\langle w\rangle$ of the work performed of a TLS as a function of the normalized driving frequency
  $\omega/\omega_0$. The left panel contains the numerical results for the fixed values $F=0.5,\,G=0.1,\,\beta=10$
  and shows a prominent maximum at $\omega_\text{max}\approx 0.941843\,\omega_0$ as well as a large number of smaller maxima. The right panel
  is devoted to the limit of small amplitudes and contains the numerical results for $F=0.05,\,G=0.01,\,\beta=10$
  (blue dots) that agree well with the analytical limit according to (\ref{W7}) (red curve). For the right panel the work is maximal at 
  $\omega_\text{max}^{(0)}\approx 0.857295 \,\omega_0$.
   }
  \label{FIGWORK}
\end{figure}

After this measurement the system evolves according to the Schr\"odinger equation (\ref{SE1}) with Hamiltonian $H(\tau)$.
At the time $\tau=2\pi$ the system hence is in the pure state $U(2\pi,0)\,P_1\, U(2\pi,0)^\ast$ with probability $\mbox{Tr}\left( P_1 W\right)$
or in the pure state $U(2\pi,0)\,P_2\, U(2\pi,0)^\ast$ with probability $\mbox{Tr}\left( P_2 W\right)$. Then a second measurement
of the instantaneous energy $H(2\pi)=H(0)$ is performed, again with the two possible outcomes $\pm \frac{\nu}{2}$. Both measurements together have
four possible outcomes symbolized by pairs $(i,j)$ where $i,j=1,2$ that occur with probabilities
\begin{equation}\label{W3}
 p_{i,j}=\mbox{Tr}\left(W\,P_i\right) \mbox{Tr}\left(P_j\,U(2\pi,0)\,P_i\, U(2\pi,0)^\ast\right)
 \;,
\end{equation}
such that $\sum_{i,j=1}^{2}p_{i,j}=1$.
The differences of the outcomes of the energy measurements yield three possible values $w=\pm \nu,\,0$ for the work
performed on the system with respective probabilities that can be calculated by using the monodromy matrix (\ref{QQ1}).
The result is identical to that obtained for the case of linear polarization in \cite{SSH20a} since it depends only on the parameters $\alpha, r$ of the
monodromy matrix.
Using the above probabilities it is straightforward to calculate the mean value of the performed work
\begin{equation}\label{W5}
  \left\langle w\right\rangle =\omega_0\left( p_{2,1}-p_{1,2}\right)=
  4\,\omega_0  r^2 \left(1-r^2\right) \sin ^2\alpha  \tanh \left(\frac{\beta  \nu    }{2}\right)\ge 0
  \;,
\end{equation}
see \cite{SSH20a}, eq.~(55). A detailed investigation of the work statistics is beyond the scope of the present article.
We will only give an example of the frequency dependence of  $\left\langle w\right\rangle$ that exhibits resonance phenomena
similar to those mentioned in Section \ref{sec:RES}, see Figure \ref{FIGWORK}.\\

However, a clear difference to the situation dealt with in Section \ref{sec:RES} is that for small amplitudes the frequency $\omega_\text{max}$
where $\left\langle w\right\rangle$ is maximal does not approach the eigenfrequency $\omega_0$ of the TLS but some other
limit $\omega_\text{max}^{(0)}$ in the interval 
\begin{equation}\label{W6}
 0.8\,\omega_0 < \omega_\text{max}^{(0)} < 0.9\, \omega_0
 \;,
\end{equation}
depending on the eccentricity of the elliptic polarization. The small amplitude limit $\left\langle w\right\rangle^{(0)}$
of $\left\langle w\right\rangle$
can be calculated by using the lowest order approximation derived in Section \ref{sec:FG} and reads:
\begin{equation}\label{W7}
\left\langle w\right\rangle^{(0)}=\frac{4\,\omega _0}{\left(\omega ^2-\omega _0^2\right)^2}\,
 \sin ^2\left(\frac{\pi  \omega _0}{\omega }\right) \tanh
   \left(\frac{\beta  \nu }{2}\right) \left(F \omega +G \omega
   _0\right)^2
   \;,
\end{equation}
see Figure \ref{FIGWORK} for an example.

%%%%%%%%%%%%%%%%%%%%%%%%%%%%%%%%%%%%%%%%%%%%%%%%%%%%%%%%%%%%%%%%%%%%%%%%%%%%%%%%%%%%%%%%%%%%%%%%%%%%%%%%%%%%%%%%%%%%%%%%%%%%%%
\section{Summary and Outlook}\label{sec:SO}
%%%%%%%%%%%%%%%%%%%%%%%%%%%%%%%%%%%%%%%%%%%%%%%%%%%%%%%%%%%%%%%%%%%%%%%%%%%%%%%%%%%%%%%%%%%%%%%%%%%%%%%%%%%%%%%%%%%%%%%%%%%%%%%%%%%%%%

The time evolution of the two level system (TLS) subject to a monochromatic, circularly polarized external field (RPC)
can be solved in terms of elementary functions, and the analogous problem with linear polarization (RPL) leads to the confluent
Heun functions. However, these two problems are only limit cases of the general
Rabi problem with elliptical polarization (RPE), and it is a natural question to
look for a solution of the latter valid in the realm where the rotating wave approximation breaks down.
This is done in the present paper by performing
the following steps:
\begin{enumerate}
  \item Reduction to the classical RPE,
  \item reduction of the classical time evolution to the first quarter period,
  \item transformation of the classical equation of motion to two $3^{rd}$ order differential equations, and
  \item solution of the latter by power series.
\end{enumerate}
This strategy has been checked by comparison with the numerical integration of the equations of motion for an example.
Moreover, we have calculated the various Fourier series of the components of the periodic solution and the
corresponding quantum or classical Floquet exponent (or quasienergy). Further, we have obtained the first terms of the power series
for the resonance frequencies w.~r.~t.~the semi-axes $F$ and $G$ of the polarization ellipse.
The latter were checked by comparison with the partially known results in the circular ($F=G$) and in the linear polarization limit ($G=0$).
This kind of result could not be obtained by a pure numerical treatment of RPE and thus justifies our analytical approach.
Analogous remarks apply to the problem of how much work is performed on a two level system by the driving field. 
For a first overview numerical methods are sufficient, see Figure \ref{FIGWORK}, but analytical methods yield 
more detailed results, e.~g., for the small amplitude limit, see Section \ref{sec:WO}.

Other limit cases that can be discussed without recourse to the $3^{rd}$ order differential equation are the adiabatic limit
($\omega\rightarrow 0$), the small amplitude limit ($F,G \rightarrow 0$) and the limit of vanishing energy splitting of the TLS
($\omega_0\rightarrow 0$). In the latter case it turns out that the exact solution of the special case $\omega_0=F=0$ cannot be transferred
to the elliptical domain except for the limit cases $F\rightarrow 0$ and $F\rightarrow G$.
Moreover, we have checked some general statements on the Rabi problem \cite{S18} like the slope relation (\ref{QIIIslope})
using our analytical approximations for some of these limit cases
as well as the power series solutions mentioned above.

It appears that this completes the set of problems related to the RPE that can be addressed with the present methods, with one exception:
In principle, it would also be possible to solve the underlying $s=1/2$ Schr\"odinger equation directly by a transformation into a third-order differential equation.  However, we have omitted this topic, firstly because of lack of space, and secondly because it is not clear which new results would follow from the direct solution.

%%%%%%%%%%%%%%%%%%%%%%%%%%%%%%%%%%%%%%%%%%%%%%%%%%%%%%%%%%%%%%%%%%%%%%%%%%%%%%%%%%%%%%%%%%%%%%%%%%%%%%%%%%%%%%%%%%%%%%%%%%%%%%%%%%%%%%%%%%
\section*{Acknowledgment}
%%%%%%%%%%%%%%%%%%%%%%%%%%%%%%%%%%%%%%%%%%%%%%%%%%%%%%%%%%%%%%%%%%%%%%%%%%%%%%%%%%%%%%%%%%%%%%%%%%%%%%%%%%%%%%%%%%%%%%%%%%%%%%%%%%%%%%%%%

I am indebted to the members of the DFG research group FOR 2692 for continuous support and encouragement, especially to Martin Holthaus and J\"urgen Schnack.  Moreover, I gratefully acknowledge discussions with Thomas Br\"ocker on the subject of this paper.

\end{document}